\documentclass[preprint,number,sort&compress]{elsarticle}

\usepackage[utf8]{inputenc}			%
\usepackage[T1]{fontenc}
\usepackage[english]{babel}
\usepackage[fleqn]{amsmath} 		%
\usepackage{amssymb}
\usepackage{mathrsfs}
\usepackage{color}
\usepackage{soul}

\usepackage{amsmath}  
\usepackage{tikz} %
\usetikzlibrary{automata,arrows,backgrounds,decorations.markings,decorations.pathmorphing,decorations.pathreplacing,fit,positioning,shadows,shapes,shapes.geometric,calc} %
\usepackage{xspace}
\usepackage[unicode,pdfmenubar,linktoc=all,hidelinks,bookmarks]{hyperref} %
\usepackage[margin=0pt,font=small,labelfont=bf]{caption} %
\usepackage{subcaption} %
\usepackage{enumerate} %
\usepackage[nomessages]{fp} %

\usepackage[plain,section]{algorithm}
\usepackage{algorithmic}

\usepackage{amsthm}
\newtheorem{theorem}{Theorem}
\newtheorem{lemma}[theorem]{Lemma}
\newtheorem{example}[theorem]{Example}
\newtheorem{corollary}[theorem]{Corollary}

\renewcommand{\algorithmicrequire}{\textbf{Input:}\@\xspace}
\renewcommand{\algorithmicensure}{\textbf{Output:}\@\xspace}

\tikzset{snake it/.style={decorate, decoration={snake,amplitude=.75mm,segment length=3mm,post length=0mm}}}
\tikzset{cut edge/.style={snake it,red,line width=2pt,font={\bfseries}}}%
\tikzset{double/.style={blue,densely dotted, very thick}}
\tikzset{inCut/.style={red, decorate, decoration={snake,amplitude=.5mm,segment length=1.5mm,post length=0mm}}}
\tikzset{inCutGray/.style={gray, decorate, decoration={snake,amplitude=.5mm,segment length=1.5mm,post length=0mm}}}
\tikzset{inCutDouble/.style={blue,densely dotted, thick, decorate, decoration={snake,amplitude=.5mm,segment length=1.5mm,post length=0mm}}}
\tikzset{edge node/.style={rectangle,minimum height=0mm,inner sep=.5mm,align=center,fill=white}}

\tikzset{nodeset one/.style={fill=green!30}}
\tikzset{nodeset two/.style={fill=blue!30}}

\tikzset{example/.style={}}
\tikzset{example node/.style={circle,inner sep=0mm,minimum size=0.8cm,thick,draw,font={\huge}}}
\tikzset{example edge/.style={draw=black,very thick}}
\tikzset{crossing edge node/.style={pos=0.4}}

\tikzset{example edge node/.style={rectangle,minimum height=4mm,inner sep=.5mm,align=center,fill=white,scale=1.4}}

\providecommand{\clipExample}{\clip (-2.6,-1.5) rectangle (4.5,5.2);}

\usepackage{xparse}

\newcommand{\sutt}[1]{_{\texttt{#1}}}
\newcommand{\ALG}[2]{\texttt{\textsc{#1}}\sutt{#2}}

\newcommand{\AlgMC}[1]{\ALG{MaxCut}{#1}}

\newcommand{\FCEMC}{\texttt{FCE-MaxCut}}

\newcommand{\MC}{\textsc{Max-Cut}}
\newcommand{\MCP}{\textsc{Max-Cut} problem\@\xspace}
\newcommand{\FCEMCA}{\textsc{Fixed\-Cut\-Edges-Max-Cut} algo\-rithm\@\xspace}
\newcommand{\MCA}{\textsc{Max-Cut} algo\-rithm\@\xspace}

\newcommand{\kfplMC}[1]{\AlgMC{#1}}

\newcommand{\kfplMCb}{\texttt{MaxCut}}

\newcommand{\contract}{\texttt{\textsc{Update}}}
\newcommand{\Split}{\texttt{\textsc{Split}}}

\newcommand{\AddEdges}{\texttt{\textsc{AddEdges}}}

\newcommand{\argmax}[1]{\underset{#1}{\text{argmax}}\;}

\newcommand{\I}{X}	%
\newcommand{\kz}{\chi}	%
\newcommand{\wert}{c}	%

\newcommand{\nodeCut}{S}	%
\newcommand{\nodeKCut}{\overline{\nodeCut}}	%
\newcommand{\nodeCutExt}[1]{\nodeCut{#1}}
\newcommand{\nodeoverlineCutExt}[1]{\overline{\nodeCut{#1}}}

\newcommand{\Cut}{\delta(\nodeCut,G)}	%
\newcommand{\CutExt}[1]{\delta(\nodeCut{#1})}

\newcommand{\etAl}{et\nolinebreak[4]\hspace{0.15em}\nolinebreak[4]al.\@\xspace}

\providecommand{\edgeAB}{1}
\providecommand{\edgeAC}{4}
\providecommand{\edgeAD}{-5}
\providecommand{\edgeAE}{-1}
\providecommand{\edgeBD}{3}
\providecommand{\edgeBE}{-3}
\providecommand{\edgeCD}{-1}
\providecommand{\edgeDE}{1}

\providecommand{\edgeCE}{2}
\providecommand{\edgeBC}{1}

\FPeval{\edgeABD}{clip(\edgeAB+\edgeAD)}
\FPeval{\edgeBDC}{clip(\edgeBC+\edgeCD)}
\FPeval{\edgeBDE}{clip(\edgeBD+\edgeBE)}
\FPeval{\edgeADE}{clip(\edgeAD+\edgeAE)}
\FPeval{\edgeCDE}{clip(\edgeCD+\edgeCE)}
\FPeval{\edgeBCE}{clip(\edgeBC+\edgeBE)}
\FPeval{\edgeCED}{clip(\edgeCD+\edgeDE)}
\FPeval{\edgeACE}{clip(\edgeAC+\edgeAE)}
\FPeval{\edgeABC}{clip(\edgeAB+\edgeAC)}
\FPeval{\edgeBCD}{clip(\edgeBD+\edgeCD)}

\FPeval{\sumOriginalCutBCE}{{clip(\edgeAB + \edgeAC + \edgeAE + \edgeBD + \edgeCD + \edgeDE)}}
\FPeval{\sumRemovedEdgeOne}{clip(\edgeAB + \edgeAC + \edgeAE + \edgeBD + \edgeCD + \edgeDE)}
\FPeval{\sumRemovedEdgeTwo}{clip(\edgeAB+\edgeAE+\edgeBC+\edgeBD+\edgeCE+\edgeDE)}
\FPeval{\sumRemovedEdgeThree}{clip(\edgeAB+\edgeAC+\edgeBD+\edgeCD+\edgeCE)}
\FPeval{\sumRemovedEdgeThreeTrueVal}{clip(\sumRemovedEdgeThree+\edgeBE)}
\FPeval{\sumMergedDE}{clip(\edgeAC+\edgeBC+\edgeCDE)}
\FPeval{\sumMergedBD}{clip(\edgeAC+\edgeBDC+\edgeCE)}

\FPeval{\negEdgeBE}{clip(-\edgeBE)}

\newcommand{\dahn}[1]{\begingroup
	#1\endgroup}
\newcommand{\kriege}[1]{\begingroup
	#1\endgroup}
\newcommand{\mutzel}[1]{\begingroup
	#1\endgroup}
\newcommand{\schilling}[1]{\begingroup
	#1\endgroup}

\begin{document}

\begin{frontmatter}

\title{Fixed-Parameter Algorithms for the \\Weighted Max-Cut Problem on Embedded 1-Planar Graphs\tnoteref{iwoca}}

\author[addr1]{Christine Dahn\corref{cor1}\fnref{fn1}}
\ead{christine.dahn@cs.uni-bonn.de}

\author[addr2]{\kriege{Nils M.~Kriege}\fnref{fn1}}
\ead{nils.kriege@univie.ac.at}

\author[addr1]{\mutzel{Petra Mutzel}\fnref{fn1}}
\ead{petra.mutzel@cs.uni-bonn.de}

\author[addr2]{\schilling{Julian Schilling}}
\ead{julian.schilling@cs.tu-dortmund.de}

\tnotetext[iwoca]{An extended abstract of this work appeared in \cite{DahnKM18}.}
\cortext[cor1]{Corresponding author}
\address[addr1]{Institute for Computer Science, University of Bonn, Bonn, Germany}
\address[addr2]{Faculty of Computer Science, University of Vienna, Vienna, Austria}
\address[addr3]{Department of Computer Science, TU Dortmund University, Dortmund, Germany}
\fntext[fn1]{This work was carried out while the authors were affiliated with TU Dortmund University.}

\begin{abstract}%

We propose two fixed-parameter tractable algorithms for the weighted \MCP on embedded 1-planar graphs 
parameterized by the crossing number $k$ of the given embedding. 
A graph is called 1-planar if it can be drawn in the plane with at most one crossing per edge.
Our algorithms recursively reduce a 1-planar graph to at most $3^k$ planar graphs, using edge removal and node contraction. 
Our main algorithm then solves the \MCP for the planar graphs using
the \FCEMC\ introduced by Liers and Pardella \cite{LiersP12}.
In the case of non-negative edge weights, we suggest a variant that allows to solve the planar instances with any planar \MCA.
We show that a maximum cut in the given 1-planar graph can be derived from the solutions for the planar graphs.
Our algorithms compute a maximum cut in an embedded weighted 1-planar graph with $n$ nodes and $k$ edge crossings in time $\mathcal{O}(3^k \cdot n^{3/2} \log n)$. %


 \end{abstract}

\end{frontmatter}

\section{Introduction}

Partitioning problems on graphs get increasing attention in the literature. Here the task is to partition the node set of a given (weighted) undirected graph, so that the number (or weighted sum) of connections between the parts is minimised.
A special case is the \MCP, which asks for a node partition into two sets, so that the sum of the edge weights in the cut is maximised.
The problem is getting increasing attention in the literature, since it is directly related to solving Ising spin glass models (see, e.g., Barahona \cite{Barahona1982}), which are of high interest in physics. Besides the theoretical merits, Ising spin glass models need to be solved in adiabatic quantum computation \cite{McGeoch2014}.
Other applications occur in the layout of electronic circuits \cite{BarahonaGJR88,DDJMRR1995}.

The \MCP has been shown to be NP-hard for general graphs \cite{Karp72}.
Papadimitriou and Yannakakis \cite{PapadimitriouY1991}  have shown that the \MCP is even APX-hard, i.e., there does not exist a polynomial-time approximation scheme, unless P=NP.
Goemans and Williamson suggested a randomized constant factor approximation algorithm \cite{GoemansW1995}, which has been derandomized by Mahajan and Ramesh \cite{MahajanR1999} and has  performance guarantee 0.87856.

In the unweighted case, i.e., all weights are one, the problem is also called the \emph{simple max-cut problem} or the \emph{maximum bipartite subgraph problem}. It is well known that in this setting always a cut of size at least $|E|/2+(|V|-1)/4$ exists that can be computed in linear time (see, e.g., \cite{NgozT93}). 
This bound is called the \emph{Edwards-Erd\H{o}s bound} and is tight for odd cliques.
Recently, Etscheid and Mnich \cite{EtscheidM18} have suggested an algorithm for the \emph{\MC{} parameterized above Edwards--Erd\H{o}s bound} problem thus improving a result by Crowston \etAl \cite{CrowstonGJ13}. This problem asks if $G$ has a cut exceeding the above bound by an amount of parameter $k$.

There are a number of special cases, for which the weighted max-cut problem can be solved in polynomial time.
The most prominent case arises if the weights of all edges are negative, since then the problem 
can be solved, e.g., via network flow.
Other special cases are, e.g., graphs without long odd cycles \cite{GrotschelN84} or weakly bipartite graphs \cite{GroetschelPulleyblank1981}.
Another prominent case appears for planar input graphs. Orlova and Dorfman \cite{OrlovaD72} and Hadlock \cite{Hadlock75} have shown that the \MCP can be solved in polynomial time for unweighted planar graphs.
Their algorithms can be extended to work on weighted planar graphs (e.g., Mutzel \cite{Mutzel90}).
The currently fastest algorithms have been suggested by Shih \etAl\ \cite{ShihWuKuo90} and by Liers and Pardella \cite{LiersP12}.
These results have been extended to the classes of graphs not contractible to $K_5$~\cite{Barahona83}  and to toroidal graphs \cite{Barahona1983,GalluccioLoebl1998}, i.e., graphs that can be embedded on the torus. %

A graph is 1-planar if it can be drawn on the plane, so that every edge is crossed at most once.
While planarity testing can be done in linear time \cite{HopcroftT1974}, the recognition problem for 1-planar graphs is much harder.
Korzhik and Mohar showed that 1-planarity testing is NP-hard \cite{KorzhikMohar13}. 
However, there are fixed-parameter tractable (FPT) algorithms parameterized by the cyclomatic number (the minimum number of edges that must be removed from the graph to make a forest), the tree-depth or the node cover number \cite{BannisterCE13}. For 1-planar graphs these algorithms construct a 1-planar embedding.

\paragraph{Our contribution}
Given an embedded weighted 1-planar graph with $k$ crossings, we introduce two fixed-parameter tractable algorithms for the \MCP with parameter $k$. The first algorithm calculates a weighted \MC\ on graphs with non-negative edge weights\footnote{This algorithm has been presented at IWOCA 2018  \cite{DahnKM18}. However, the claim that the algorithm works for general edge-weights is not correct, as shown in this paper.} 
and the second on graphs with arbitrary edge weights. The main idea of our algorithms is to recursively reduce the input graph into a set of at most $3^k$ planar graphs using a series of edge removals and node contractions. The planar instances can then be solved using the polynomial time algorithm suggested in \cite{LiersP12} with running time $O(n^{3/2}\log n)$ for a planar graph with $n$ nodes.

The paper is organized as follows. Section \ref{se:preliminaries} contains the basic definitions concerning cuts and 1-planarity. 
We also introduce the class of $k$-almost-planar graphs, which admit a 1-planar drawing not exceeding $k$ crossings.
In Section \ref{se:alg-1pl} we present our weighted \MCA{} for embedded 1-planar graphs with non-negative edge weights and prove its correctness. In Section \ref{se:co-ex-neg-w} we give a counter example to show the first algorithm might fail on graphs with negative edge weights. Therefore a variant of our algorithm is introduced in Section \ref{se:alg-neg}. The second algorithm is proven to solve the weighted \MCP{} on embedded graphs with arbitrary edge weights.
Our analysis of the running times shows that both algorithms are fixed-parameter tractable with parameter $k$. 
We end with a conclusion and open problems in Section \ref{se:conclusion}.

\dahn{
\paragraph{Follow-up work}

In follow-up works Chimani \etAl \cite{CDJKMN19} and Kobayashi \etAl \cite{KKMT19I} 
extended our algorithm to solve the weighted \MCP on arbitrary graphs. 
Using different techniques, both results improve the running time of the algorithm to 
$\mathcal{O}(2^k (n+k)^{3/2}\log (n+k))$, where $n$ is the number of nodes and $k$ is 
the number of crossings in a given embedded graph.
Furthermore Chimani \etAl \cite{CDJKMN19} show that \MC\ is fixed-parameter tractable with respect to\ the crossing number, even without a given embedding.
} 
\section{Preliminaries}
\label{se:preliminaries}

Throughout our paper we consider undirected weighted graphs $G=(V,E,c)$ with non-negative edge weights in Section \ref{se:alg-1pl} and arbitrary edge weights in Sections \ref{se:co-ex-neg-w} and \ref{se:alg-neg}.
A partition of the nodes of $G$ into two sets $S \subseteq V$ and $\overline{S}=V\backslash S$
defines the \emph{cut} $F=\Cut=\{uv \in E \mid (u\in S \hbox{ and }v\in \overline{S}) \hbox{ or } (v\in S \hbox{ and }u\in \overline{S})  \}$.
The \emph{value of a cut} $F$ in the graph $G$ is the sum of weights of all edges in the cut:
$\wert(F)= \sum_{e\in F} c_{e}$. 
Choosing $S=\emptyset$ gives a valid cut of value $0$.
The weighted \MCP searches for a cut in a given weighted graph with highest value.
For the graph class of planar graphs, the \MCP can be solved in polynomial time \cite{Hadlock75,OrlovaD72}.

A graph is \emph{planar} if it admits a \emph{planar drawing}, i.e., a drawing on the plane without any edge crossing.
A drawing admits a \emph{rotation system} $\Pi$, which is a clockwise-ordering of the incident edges for every node. In a planar drawing, a rotation system defines the \emph{faces}, i.e., the topologically connected regions of the plane.
One of the faces, the \emph{outer face}, is unbounded. 
A face is uniquely described by its boundary edges. 
Such a description for each face is an equivalent definition of a (planar) embedding. 
A \emph{(planar) embedding} represents the set of all planar drawings with the same faces. It can be represented by the description of the faces or by the rotation system.
It is well known that planarity testing can be solved in linear time \cite{HopcroftT73}. The same is true for computing a planar embedding~\cite{MehlhornM1996}.
In order to generate crossing free drawings of planar graphs, a number of various algorithms exist, e.g., the straight-line drawing algorithm by de Fraysseix et al.~\cite{DeFraysseixPP1990}.

Planar graphs are contained in the class of 1-planar graphs. A graph is \emph{1-planar} if it admits a \emph{1-planar drawing}, i.e., a drawing on the plane with at most one crossing per edge.
Testing 1-planarity is NP-hard \cite{KorzhikMohar13} even in the case of bounded treewidth or bandwith \cite{BannisterCE13}. 
A \emph{1-planar embedding} defines the faces of a given 1-planar drawing, i.e., the topologically connected regions of the plane, and can be represented by the set of crossings $X \subset 2^E$ and a rotation system $\Pi$ for the nodes of the graph. 
A face of a 1-planar drawing is uniquely described by its boundary edges and \textit{half edges}, i.e., an edge between a node and a crossing.
Note that a 1-planar embedding uniquely defines a rotation system for the nodes. 
However, the opposite is not true. In general, a rotation system does not allow for computing the crossings or a 1-planar embedding efficiently.
Auer et al. \cite{AuerBGR15} have shown that testing 1-planarity of a graph with a fixed rotation system is NP-hard even if the graph is 3-connected.

We call a 1-planar graph \emph{$k$-almost-planar} if it admits a 1-planar drawing with at most $k$ edge crossings.
For edge removal and node contraction we use the following notation: $G - e = (V,E\setminus\{e\})$ denotes the graph obtained from $G=(V,E)$ by deleting the edge $e\in E$. $G/xy$ denotes the graph obtained by contracting the two nodes $x$ and $y$ into a new node $v_{xy}\notin V$. In doing so the edges leading to $x$ or $y$ are replaced by a new edge to $v_{xy}$. Multi-edges to $v_{xy}$ are contracted to one edge and their edge weights are added, self-loops are deleted.
We denote the inverse operation of contraction by $\Split$.
The contraction and $\Split$ operation can be applied to a subset of nodes $S\subseteq V$ as well:

\begin{align*}
S/xy = & \begin{cases}
S\setminus\{x,y\} \cup \{v_{xy}\} & \text{ if } x,y \in S\\
S &  \text{ otherwise} 
\end{cases}\\
\Split(S,v_{xy}) = & \begin{cases}
S\setminus\{v_{xy}\} \cup \{x,y\} & \text{ if } v_{xy} \in S\\
S & \text{ otherwise}
\end{cases}
\end{align*}

\section{Max-Cut for embedded 1-planar graphs with non-negative edge weights}
\label{se:alg-1pl}

Our main idea for computing the maximum cut in an embedded weighted 1-planar graph with non-negative edge weights is to eliminate its $k$ crossings and then use a \MCA for planar graphs on the resulting planar graphs. In order to remove a crossing we need to know its two crossing edges. 
We introduce two methods to remove a crossing: Either by deleting one of the crossing edges, or by contracting two nodes that do not belong to the same crossing edge.

\subsection{Removing a crossing}
\label{suse:alg-1pl:remCross}
In this section let $G=(V,E,c)$ be a $k$-almost-planar graph with a 1-planar embedding $(\I,\Pi)$ and a set of crossing edges $\I$ with $\vert \I \vert = k$.
A crossing is defined by a pair of crossing edges, e.g., let $\kz = \{e_{vy}, e_{wz}\} \in \I$ be an arbitrary crossing. 
The following lemma shows that specific node contractions (and edge deletions) remove at least one crossing and do not introduce new crossings.

\begin{lemma}
	\label{lem:k-1-al-pl}
	Let $G$ be a $k$-almost-planar graph with a 1-planar embedding $(\I,\Pi)$ and let $\kz = \{e_{vy}, e_{wz}\} \in \I$ be an arbitrary crossing.
	The graphs $G/ab$, $G - e_{vy}$ and $G - e_{wz}$ are $(k-1)$-almost-planar for $ab \in \{vw,vz,wy,yz\}$.
	The set of crossings in each resulting 1-planar embedding is a proper subset of $\I$.
\end{lemma}
\begin{proof}
	Since the contracted nodes $a$ and $b$ are each an endpoint of one of the crossing edges, the contracted node is an endpoint to both edges. Since $e_{vy}$ and $e_{wz}$ now have a common endpoint, they can be drawn in the plane without crossing. 
	Therefore the crossing $\kz$ is removed. 
	The contraction does not create new crossings because the two nodes $a$ and $b$ can be moved along their half edges  towards the crossing. 
	This is possible because in every a 1-planar embedding every crossing is incident to four half edges connecting it with its four endpoints. 
	The new node $v_{ab}$ is then placed where the crossing used to be. All other edges can be extended to the new node along the way of the same half edges without creating new crossings. Multi-edges are merged into one of the two edges and self-loops are deleted.
	In $G - e_{vy}$ and $G - e_{wz}$ the crossing $\kz$ is removed by deleting one of its crossing edges.
	Obviously this does not lead to new crossings.
	So in both cases the number of crossings decreases.
\end{proof}

\begin{figure}[tb]
	\centering
	\begin{subfigure}{.49\linewidth}
		\centering
		\scalebox{0.9}{%
\begin{tikzpicture}
	\node[draw,circle,minimum size=0.45cm, inner sep=0ex] (a) at (0,0) {\small $a$};
	\node[draw,circle,minimum size=0.45cm, inner sep=0ex] (b) at (2,0) {\small $b$};
	\node[draw,circle,minimum size=0.45cm, inner sep=0ex] (c) at (2,2) {\small $c$};
	\node[draw,circle,minimum size=0.45cm, inner sep=0ex] (d) at (0,2) {\small $d$};
	\node[draw,circle,minimum size=0.45cm, inner sep=0ex] (e) at (4,0) {\small $e$};
	\node[draw,circle,minimum size=0.45cm, inner sep=0ex] (f) at (4,2) {\small $f$};

	\node[draw,circle,minimum size=0.45cm, inner sep=0ex] (ab) at (1,0) {\small $t$};
	\node[draw,circle,minimum size=0.45cm, inner sep=0ex] (cd) at (1,2) {\small $s$};
	\node[draw,circle,minimum size=0.45cm, inner sep=0ex] (be) at (3,0) {\small $y$};
	\node[draw,circle,minimum size=0.45cm, inner sep=0ex] (cf) at (3,2) {\small $x$};

	\draw (a) -- (ab);
	\draw (ab) -- (b);
	\draw (a) -- (c);
	\draw (a) -- (d);
	\draw (b) -- (c);
	\draw (b) -- (d);
	\draw (c) -- (cd);
	\draw (cd) -- (d);
	\draw (b) -- (be);
	\draw (be) -- (e);
	\draw (b) -- (f);
	\draw (c) -- (e);
	\draw (c) -- (cf);
	\draw (cf) -- (f);
	\draw (e) -- (f);

	\draw[-] (ab) to [out=-140,in=-90,looseness=1.5] (-1,1);
	\draw[-] (cd) to [out=140,in=90,looseness=1.5] (-1,1);
	\draw[-] (be) to [out=-50,in=-90,looseness=1.5] (5,1);
	\draw[-] (cf) to [out=50,in=90,looseness=1.5] (5,1);
\end{tikzpicture} %
}
		\caption{$H$}
		\label{fig:rm_cr_0}
	\end{subfigure}
	\begin{subfigure}{.49\linewidth}
		\centering
		\scalebox{0.9}{%
\begin{tikzpicture}
	\node[draw,circle,minimum size=0.45cm, inner sep=0ex] (a) at (0,0) {\small $a$};
	\node[draw,circle,minimum size=0.45cm, inner sep=0ex] (vbc) at (1,1) {\small $v_{bc}$};
	\node[draw,circle,minimum size=0.45cm, inner sep=0ex]  (d) at (0,2) {\small $d$};
	\node[draw,circle,minimum size=0.45cm, inner sep=0ex] (e) at (4,0) {\small $e$};
	\node[draw,circle,minimum size=0.45cm, inner sep=0ex]  (f) at (4,2) {\small $f$};

	\node[draw,circle,minimum size=0.45cm, inner sep=0ex]  (ab) at (1,0) {\small $t$};
	\node[draw,circle,minimum size=0.45cm, inner sep=0ex] (cd) at (1,2) {\small $s$};
	\node[draw,circle,minimum size=0.45cm, inner sep=0ex]  (be) at (3,0) {\small $y$};
	\node[draw,circle,minimum size=0.45cm, inner sep=0ex] (cf) at (3,2) {\small $x$};

	\draw (a) -- (ab);
	\draw (ab) -- (vbc);
	\draw (a) -- (vbc);
	\draw (a) -- (d);
	\draw (vbc) -- (d);
	\draw (vbc) -- (cd);
	\draw (cd) -- (d);
	\draw (vbc) -- (2,0)-- (be);
	\draw (be) -- (e);
	\draw (vbc) -- (3,1.25) -- (f);
	\draw (vbc) -- (3,0.75) -- (e);
	\draw (vbc) -- (2,2)-- (cf);
	\draw (cf) -- (f);
	\draw (e) -- (f);

	\draw[-,] (ab) to [out=-140,in=-90,looseness=1.5] (-1,1) to [out=90,in=140,looseness=1.5] (cd);
	\draw[-,] (be) to [out=-50,in=-90,looseness=1.5] (5,1) to [out=90,in=50,looseness=1.5] (cf);
\end{tikzpicture}
\vspace*{-6pt} %
}
		\caption{$H/bc$}
		\label{fig:rm_cr_4}
	\end{subfigure}
	\begin{subfigure}{.49\linewidth}
		\centering
		\scalebox{0.9}{%
\begin{tikzpicture}
	\node[draw,circle,minimum size=0.45cm, inner sep=0ex] (a) at (0,0) {\small $a$};
	\node[draw,circle,minimum size=0.45cm, inner sep=0ex] (b) at (2,0) {\small $b$};
	\node[draw,circle,minimum size=0.45cm, inner sep=0ex]  (vcd) at (1,1) {\small $v_{cd}$};
	\node[draw,circle,minimum size=0.45cm, inner sep=0ex] (e) at (4,0) {\small $e$};
	\node[draw,circle,minimum size=0.45cm, inner sep=0ex]  (f) at (4,2) {\small $f$};

	\node[draw,circle,minimum size=0.45cm, inner sep=0ex]  (ab) at (1,0) {\small $t$};
	\node[draw,circle,minimum size=0.45cm, inner sep=0ex] (cd) at (1,2) {\small $s$};
	\node[draw,circle,minimum size=0.45cm, inner sep=0ex]  (be) at (3,0) {\small $y$};
	\node[draw,circle,minimum size=0.45cm, inner sep=0ex] (cf) at (3,2) {\small $x$};

	\draw (a) -- (ab);
	\draw (ab) -- (b);
	\draw (a) edge node[edge node] {2} (vcd);
	\draw (b) edge node[edge node] {2} (vcd);
	\draw (vcd) edge node[left=0.5mm,edge node] {2} (cd);
	
	\draw (b) -- (be);
	\draw (be) -- (e);
	\draw (b) -- (f);
	\draw (vcd) -- (2.25,1.75) -- (e);
	\draw (vcd) -- (2,2) -- (cf);
	\draw (cf) -- (f);
	\draw (e) -- (f);

	\draw[-] (ab) to [out=-140,in=-90,looseness=1.5] (-1,1) to [out=90,in=140,looseness=1.5] (cd);
	\draw[-] (be) to [out=-50,in=-90,looseness=1.5] (5,1) to [out=90,in=50,looseness=1.5] (cf);
\end{tikzpicture}
\vspace*{-6pt} %
}
		\caption{$H/cd$}
		\label{fig:rm_cr_3}
	\end{subfigure}
	\begin{subfigure}{.49\linewidth}
		\centering
		\scalebox{0.9}{%
\begin{tikzpicture}
	\node[draw,circle,minimum size=0.45cm, inner sep=0ex] (a) at (0,0) {\small $a$};
	\node[draw,circle,minimum size=0.45cm, inner sep=0ex] (b) at (2,0) {\small $b$};
	\node[draw,circle,minimum size=0.45cm, inner sep=0ex]  (c) at (2,2) {\small $c$};
	\node[draw,circle,minimum size=0.45cm, inner sep=0ex]  (d) at (0,2) {\small $d$};
	\node[draw,circle,minimum size=0.45cm, inner sep=0ex] (e) at (4,0) {\small $e$};
	\node[draw,circle,minimum size=0.45cm, inner sep=0ex]  (f) at (4,2) {\small $f$};

	\node[draw,circle,minimum size=0.45cm, inner sep=0ex]  (ab) at (1,0) {\small $t$};
	\node[draw,circle,minimum size=0.45cm, inner sep=0ex] (cd) at (1,2) {\small $s$};
	\node[draw,circle,minimum size=0.45cm, inner sep=0ex]  (be) at (3,0) {\small $y$};
	\node[draw,circle,minimum size=0.45cm, inner sep=0ex] (cf) at (3,2) {\small $x$};

	\draw (a) -- (ab);
	\draw (ab) -- (b);
	\draw (a) -- (c);
	\draw (a) -- (d);
	\draw (b) -- (c);
	\draw (c) -- (cd);
	\draw (cd) -- (d);
	\draw (b) -- (be);
	\draw (be) -- (e);
	\draw (b) -- (f);
	\draw (c) -- (e);
	\draw (c) -- (cf);
	\draw (cf) -- (f);
	\draw (e) -- (f);

	\draw[-] (ab) to [out=-140,in=-90,looseness=1.5] (-1,1) to [out=90,in=140,looseness=1.5] (cd);
	\draw[-] (be) to [out=-50,in=-90,looseness=1.5] (5,1) to [out=90,in=50,looseness=1.5] (cf);
\end{tikzpicture}
\vspace*{-6pt} %
}
		\caption{$H - e_{bd}$}
		\label{fig:rm_cr_2}
	\end{subfigure}
	\caption{An example of how a crossing can be removed in three different ways. (All edges have weight $1$, except the merged edges in $H/cd$. They have weight $2$.)}
	\label{fig:rm_cr}
\end{figure}
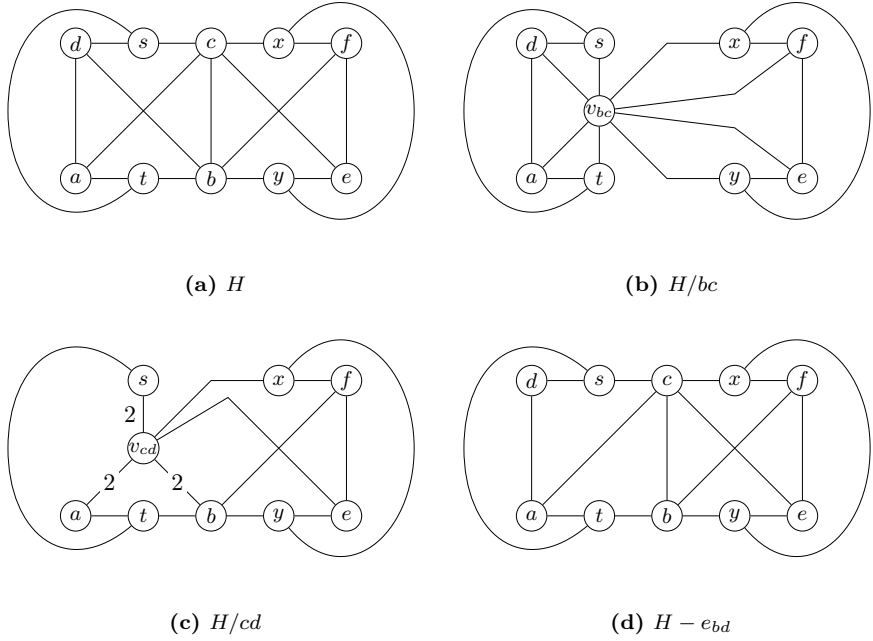

Two examples for node contraction are given in Figures~\ref{fig:rm_cr_4} and~\ref{fig:rm_cr_3}. Note that the contraction shown in Figure \ref{fig:rm_cr_4} removed both crossings of $H$, cf. Figure \ref{fig:rm_cr_0}. An example for edge deletion is given in Figure \ref{fig:rm_cr_2}.
The recursive application of Lemma \ref{lem:k-1-al-pl} shows that all crossings can be removed with these two operations. Thus after $k$ contraction or removal operations the resulting graph is planar and a planar \MCA can be applied to compute a maximum cut.

The following lemma shows how to project a cut in $G/ab$ or $G - e_{ab}$ back onto $G$. Note that we require $a$ and $b$ to be on the same side of the given cut in $G-e_{ab}$ for the cut to have the same value in $G - e_{ab}$ and $G$.

\begin{lemma}
	\label{lem:cut-transfer-to-G}\ 
	Let $G=(V,E,c)$ be a weighted graph, $a, b\in V$ and $a \neq b$.%
	\begin{enumerate}[(i)]
		\item If $\CutExt{,G/ab}$ is a cut in $G/ab$, then the cut $\delta(\Split(\nodeCut,v_{ab}),G)$ has the same value in $G$ as $\CutExt{,G/ab}$ in $G/ab$. 
		\label{lem:cut-transfer-to-G_contraction}
		\item 
		If $a$ and $b$ are on the same side of the cut $\CutExt{,G - e_{ab}}$, i.e. $a,b \in \nodeCut$ or $a,b \in \nodeKCut$ for $S\subseteq V$, 
		then $\Cut = \CutExt{,G - e_{ab}}$ for $e_{ab}\in E$.
		\label{lem:cut-transfer-to-G_removal}
	\end{enumerate}
\end{lemma}
\begin{proof}
	\textit{(\ref{lem:cut-transfer-to-G_contraction})} For $\nodeCut\subseteq V\setminus\{a,b\} \cup \{v_{ab}\}$ the set $\nodeCut$ defines a cut in $G/ab$. If the contracted node $v_{ab}$ is split the cut is projected from $G/ab$ to $G$. The corresponding set of nodes in $G$ is $\nodeCutExt{'}=\Split(\nodeCut,v_{ab})$. It defines a cut in $G$. If $v_{ab} \notin \nodeCut$ then $\nodeCutExt{'} = \nodeCut$.
	The weight of an edge $e \in \CutExt{,G/ab}$ in $G/ab$ is either the same as the weight of the corresponding edge $e' \in \CutExt{',G}$ in $G$ or it is split between two edges $e', e'' \in \CutExt{',G}$ in $G$. 
	The only edge that might exist in $G$ but not in $G/ab$ is $e_{ab}$. %
	Since $a$ and $b$ were contracted %
	in $G/ab$, they are %
	either both in $\nodeCutExt{'}$ or both in $\nodeoverlineCutExt{'}$. 
	Therefore the only edge 
	that could be added in $G$ by splitting $v_{ab}$ 
	can not add to the value of $\CutExt{',G}$ in $G$.
	So no weights are lost or added due to the projection and the two cuts have the same value.\\
	\textit{(\ref{lem:cut-transfer-to-G_removal})} This is obvious because $e_{ab}$ is in neither of the two cuts.
\end{proof}

\subsection{The Algorithm}
\label{suse:alg-1pl:algo}
We use three operations to successively remove all 
crossings of a 1-planar graph. All planar instances obtained in this way are
then solved by a \MCA for planar graphs. From the solutions of the 
planar graphs a solution for the original graph is constructed.
Note that the algorithm does only need the graph $G$ and the set of edge crossings $\I$ as input.
However, the 1-planar embedding is needed to show the correctness of the algorithm.

\begin{algorithm}[tb]

\underline{$\kfplMCb(G,\I)$}\vspace{0.5em}\\
\algorithmicrequire \ %
An undirected weighted 1-planar graph $G$ with non-negative edge weigths and a set of crossing edges $\I$ in a 1-planar embedding of $G$. \\
\algorithmicensure \ A set $\nodeCut \subseteq V_G$ defining a maximum cut $\Cut \subseteq E_G$ in $G$.

\begin{algorithmic}[1]
\IF{$\I = \emptyset$}
\STATE $\nodeCut \gets \kfplMC{planar}(G)$  									\label{algo:MaxCut1pl-old:plMC}
\ELSE
\STATE choose an element $\kz\gets\{e_{vy}, e_{wz}\} \in \I$
\STATE $G_1 \gets G/wy$
\STATE $\nodeCutExt{_1} \gets \kfplMCb(G_1, \contract(\I,w,y))$				\label{algo:MaxCut1pl-old:f1}
\STATE $G_2 \gets G/yz$
\STATE $\nodeCutExt{_2} \gets \kfplMCb(G_2, \contract(\I,y,z))$				\label{algo:MaxCut1pl-old:f2}
\STATE $G_3 \gets G-e_{wz}$
\STATE $\nodeCutExt{_3} \gets \kfplMCb(G_3,\I \setminus \{\kz\})$			\label{algo:MaxCut1pl-old:f3}
\STATE $j \gets \argmax{1\leq i\leq 3} \wert(\CutExt{_i,G_i})$					\label{algo:MaxCut1pl-old:maxGew}
\IF{$j = 1$}																	\label{algo:MaxCut1pl-old:convert_S}
\STATE $\nodeCut \gets \Split(\nodeCutExt{_1},v_{wy})$							\label{algo:MaxCut1pl-old:KS1}
\ELSIF{$j = 2$}
\STATE $\nodeCut \gets \Split(\nodeCutExt{_2},v_{yz})$							\label{algo:MaxCut1pl-old:KS2}	
\ELSE
\STATE $\nodeCut \gets \nodeCutExt{_3}$
\ENDIF																			\label{algo:MaxCut1pl-old:convert_E}
\ENDIF
\STATE \textbf{return} $\nodeCut$												\label{algo:MaxCut1pl-old:return}
\end{algorithmic}
\caption{Weighted \MCA for embedded 1-planar graphs with non-negative edge weigths.}

 	\label{algo:MaxCut1pl}
\end{algorithm}

Algorithm \ref{algo:MaxCut1pl} realizes this approach by a recursive function,
which is initially called with the input graph $G$ and the set of crossings $\I$
present in its embedding.
As the algorithm progresses, the graph is successively modified and the set of
crossings is adjusted according to the modifications applied.
If the graph $G$, passed as parameter to the function, is planar ($X=\emptyset$), 
then a planar \MCA is called (line~\ref{algo:MaxCut1pl-old:plMC}). 
If there are still crossings remaining, an arbitrary crossing is selected and 
removed in three different ways: 
Let $y$ be an arbitrary endpoint of one crossing edge and $e_{wz}$, $w \neq y$,
$z \neq y$, the other crossing edge, then (I) the nodes $y$ and $w$ are contracted, 
(II) the nodes $y$ and $z$ are contracted, and (III) the edge $e_{wz}$ is deleted.
Each operation removes at least the selected crossing, but in cases (I) and (II) 
other crossing may be affected as well. Therefore, the set of crossings $\I$ is 
adjusted by the function $\contract$.
If two nodes $w,y$ are contracted, $\contract(\I,w,y)$ removes every crossing in $\I$, 
which was dissolved by contracting $w$ and $y$, and replaces every appearance of $w$ 
or $y$ in $\I$ with the contracted node $v_{wy}$. To check if a crossing was dissolved, 
$\contract$ checks if $w$ and $y$ are both part of the crossing. Since every crossing 
needs to be checked once, $\contract$ has a linear running time.
For each case the recursive function is called with the modified graph and crossing
set as a parameter (lines~\ref{algo:MaxCut1pl-old:f1}-\ref{algo:MaxCut1pl-old:f3}).
Each call returns a node set defining a maximum cut in the modified instance.
The cut with the maximum value is then transferred back to $G$. If the maximum cut
is obtained for a graph with contracted nodes, i.e., case (I) or (II), then the
original nodes are restored by the function $\Split(\nodeCut_i,v_{wy})$, which 
replaces $v_{wy}$ with $w$ and $y$ if $\nodeCut_i$ contains the contracted node.
If the maximum cut is obtained for a graph with a deleted edge, i.e., case (III), 
the nodeset is passed directly to the return value $S$. In this case we can assume 
that the two nodes of the removed edge are on the same side of the cut. Otherwise 
one of the other two cuts would have an equal value and would have been chosen 
instead.
The cut defining set $S$ is then returned as the solution to the subproblem.

\begin{figure}[tb]
	\centering
	\begin{subfigure}{.44\linewidth}
		\centering
		\scalebox{0.9}{%
\begin{tikzpicture}
	\node[draw,circle,minimum size=0.45cm, inner sep=0ex, nodeset two] (a) at (0,0) {\small $a$};
	\node[draw,circle,minimum size=0.45cm, inner sep=0ex, nodeset two] (vbc) at (1,1) {\small $v_{bc}$};
	\node[draw,circle,minimum size=0.45cm, inner sep=0ex, nodeset one]  (d) at (0,2) {\small $d$};
	\node[draw,circle,minimum size=0.45cm, inner sep=0ex, nodeset two] (e) at (4,0) {\small $e$};
	\node[draw,circle,minimum size=0.45cm, inner sep=0ex, nodeset one]  (f) at (4,2) {\small $f$};

	\node[draw,circle,minimum size=0.45cm, inner sep=0ex, nodeset one]  (ab) at (1,0) {\small $t$};
	\node[draw,circle,minimum size=0.45cm, inner sep=0ex, nodeset two] (cd) at (1,2) {\small $s$};
	\node[draw,circle,minimum size=0.45cm, inner sep=0ex, nodeset one]  (be) at (3,0) {\small $y$};
	\node[draw,circle,minimum size=0.45cm, inner sep=0ex, nodeset two] (cf) at (3,2) {\small $x$};

	\draw[inCut] (a) -- (ab);
	\draw[inCut] (ab) -- (vbc);
	\draw[] (a) -- (vbc);
	\draw[inCut] (a) -- (d);
	\draw[inCut] (vbc) -- (d);
	\draw[] (vbc) -- (cd);
	\draw[inCut] (cd) -- (d);
	\draw[inCut] (vbc) -- (2,0)-- (be);
	\draw[inCut] (be) -- (e);
	\draw[inCut] (vbc) -- (3,1.25) -- (f);
	\draw[] (vbc) -- (3,0.75) -- (e);
	\draw[] (vbc) -- (2,2)-- (cf);
	\draw[inCut] (cf) -- (f);
	\draw[inCut] (e) -- (f);

	\draw[-,inCut] (ab) to [out=-140,in=-90,looseness=1.5] (-1,1) to [out=90,in=140,looseness=1.5] (cd);
	\draw[-,inCut] (be) to [out=-50,in=-90,looseness=1.5] (5,1) to [out=90,in=50,looseness=1.5] (cf);
\end{tikzpicture}
\vspace*{-6pt} %
}
		\caption{\MC\ in $H/bc$}
		\label{fig:kfplMC_4}
	\end{subfigure}	
	\begin{subfigure}{.44\linewidth}
		\centering
		\scalebox{0.9}{%
\begin{tikzpicture}
	\node[draw,circle,minimum size=0.45cm, inner sep=0ex, nodeset two] (a) at (0,0) {\small $a$};
	\node[draw,circle,minimum size=0.45cm, inner sep=0ex, nodeset two] (b) at (2,0) {\small $b$};
	\node[draw,circle,minimum size=0.45cm, inner sep=0ex, nodeset one]  (vcd) at (1,1) {\small $v_{cd}$};
	\node[draw,circle,minimum size=0.45cm, inner sep=0ex, nodeset two] (e) at (4,0) {\small $e$};
	\node[draw,circle,minimum size=0.45cm, inner sep=0ex, nodeset one]  (f) at (4,2) {\small $f$};

	\node[draw,circle,minimum size=0.45cm, inner sep=0ex, nodeset one]  (ab) at (1,0) {\small $t$};
	\node[draw,circle,minimum size=0.45cm, inner sep=0ex, nodeset two] (cd) at (1,2) {\small $s$};
	\node[draw,circle,minimum size=0.45cm, inner sep=0ex, nodeset one]  (be) at (3,0) {\small $y$};
	\node[draw,circle,minimum size=0.45cm, inner sep=0ex, nodeset two] (cf) at (3,2) {\small $x$};

	\draw[inCut] (a) -- (ab);
	\draw[inCut] (ab) -- (b);
	\draw (a) edge[inCut] node[edge node] {2} (vcd);
	
	\draw [-] (b) edge[inCut] node[edge node] {2} (vcd);
	
	\draw (vcd) edge[inCut] node[left=0.5mm,edge node] {2} (cd);
	\draw[inCut] (b) -- (be);
	\draw[inCut] (be) -- (e);
	\draw[inCut] (b) -- (f);
	\draw[inCut] (vcd) -- (2.25,1.75) -- (e);
	\draw[inCut] (vcd) -- (2,2) -- (cf);
	\draw[inCut] (cf) -- (f);
	\draw[inCut] (e) -- (f);

	\draw[-,inCut] (ab) to [out=-140,in=-90,looseness=1.5] (-1,1) to [out=90,in=140,looseness=1.5] (cd);
	\draw[-,inCut] (be) to [out=-50,in=-90,looseness=1.5] (5,1) to [out=90,in=50,looseness=1.5] (cf);
\end{tikzpicture}
\vspace*{-6pt} %
}		
		\caption{\MC\ in $H/cd$}
		\label{fig:kfplMC_3}
	\end{subfigure}	
	\begin{subfigure}{.44\linewidth}
		\centering
		\scalebox{0.9}{%
\begin{tikzpicture}
	\node[draw,circle,minimum size=0.45cm, inner sep=0ex, nodeset two] (a) at (0,0) {\small $a$};
	\node[draw,circle,minimum size=0.45cm, inner sep=0ex, nodeset two] (b) at (2,0) {\small $b$};
	\node[draw,circle,minimum size=0.45cm, inner sep=0ex, nodeset one]  (c) at (2,2) {\small $c$};
	\node[draw,circle,minimum size=0.45cm, inner sep=0ex, nodeset one]  (d) at (0,2) {\small $d$};
	\node[draw,circle,minimum size=0.45cm, inner sep=0ex, nodeset two] (e) at (4,0) {\small $e$};
	\node[draw,circle,minimum size=0.45cm, inner sep=0ex, nodeset one]  (f) at (4,2) {\small $f$};

	\node[draw,circle,minimum size=0.45cm, inner sep=0ex, nodeset one]  (ab) at (1,0) {\small $t$};
	\node[draw,circle,minimum size=0.45cm, inner sep=0ex, nodeset two] (cd) at (1,2) {\small $s$};
	\node[draw,circle,minimum size=0.45cm, inner sep=0ex, nodeset one]  (be) at (3,0) {\small $y$};
	\node[draw,circle,minimum size=0.45cm, inner sep=0ex, nodeset two] (cf) at (3,2) {\small $x$};

	\draw[inCut] (a) -- (ab);
	\draw[inCut] (ab) -- (b);
	\draw[inCut] (a) -- (c);
	\draw[inCut] (a) -- (d);
	\draw[inCut] (b) -- (c);
	\draw[inCut] (c) -- (cd);
	\draw[inCut] (cd) -- (d);
	\draw[inCut] (b) -- (be);
	\draw[inCut] (be) -- (e);
	\draw[inCut] (b) -- (f);
	\draw[inCut] (c) -- (e);
	\draw[inCut] (c) -- (cf);
	\draw[inCut] (cf) -- (f);
	\draw[inCut] (e) -- (f);

	\draw[-,inCut] (ab) to [out=-140,in=-90,looseness=1.5] (-1,1) to [out=90,in=140,looseness=1.5] (cd);
	\draw[-,inCut] (be) to [out=-50,in=-90,looseness=1.5] (5,1) to [out=90,in=50,looseness=1.5] (cf);
\end{tikzpicture}
\vspace*{-6pt} %
}
		\caption{\MC\ in $H - e_{db}$}
		\label{fig:kfplMC_2}
	\end{subfigure}	
	\begin{subfigure}{.44\linewidth}
		\centering
		\scalebox{0.9}{%
\begin{tikzpicture}
	\node[draw,circle,minimum size=0.45cm, inner sep=0ex, nodeset two] (a) at (0,0) {\small $a$};
	\node[draw,circle,minimum size=0.45cm, inner sep=0ex, nodeset two] (b) at (2,0) {\small $b$};
	\node[draw,circle,minimum size=0.45cm, inner sep=0ex, nodeset one]  (c) at (2,2) {\small $c$};
	\node[draw,circle,minimum size=0.45cm, inner sep=0ex, nodeset one]  (d) at (0,2) {\small $d$};
	\node[draw,circle,minimum size=0.45cm, inner sep=0ex, nodeset two] (e) at (4,0) {\small $e$};
	\node[draw,circle,minimum size=0.45cm, inner sep=0ex, nodeset one]  (f) at (4,2) {\small $f$};

	\node[draw,circle,minimum size=0.45cm, inner sep=0ex, nodeset one]  (ab) at (1,0) {\small $t$};
	\node[draw,circle,minimum size=0.45cm, inner sep=0ex, nodeset two] (cd) at (1,2) {\small $s$};
	\node[draw,circle,minimum size=0.45cm, inner sep=0ex, nodeset one]  (be) at (3,0) {\small $y$};
	\node[draw,circle,minimum size=0.45cm, inner sep=0ex, nodeset two] (cf) at (3,2) {\small $x$};

	\draw[inCut] (a) -- (ab);
	\draw[inCut] (ab) -- (b);
	\draw[inCut] (a) -- (c);
	\draw[inCut] (a) -- (d);
	\draw[inCut] (b) -- (c);
	\draw[inCut] (b) -- (d);
	\draw[inCut] (c) -- (cd);
	\draw[inCut] (cd) -- (d);
	\draw[inCut] (b) -- (be);
	\draw[inCut] (be) -- (e);
	\draw[inCut] (b) -- (f);
	\draw[inCut] (c) -- (e);
	\draw[inCut] (c) -- (cf);
	\draw[inCut] (cf) -- (f);
	\draw[inCut] (e) -- (f);

	\draw[-,inCut] (ab) to [out=-140,in=-90,looseness=1.5] (-1,1) to [out=90,in=140,looseness=1.5] (cd);
	\draw[-,inCut] (be) to [out=-50,in=-90,looseness=1.5] (5,1) to [out=90,in=50,looseness=1.5] (cf);
\end{tikzpicture}
\vspace*{-6pt} %
}
		\caption{\MC\ in $H$}
		\label{fig:kfplMC_MC}
	\end{subfigure}
	\caption{An example how the algorithm calculates a \MC~in an embedded 2-almost-
		planar graph. (Merged edges have weight 2; all other edges have weight 1. 
		Red-curvy edges belong to the cut; black-straight edges do not belong to 
		the cut. The color of the nodes indicates on which side of the partition, 
		defining the cut, a node is.)}
	\label{fig:bsp5}
\end{figure}
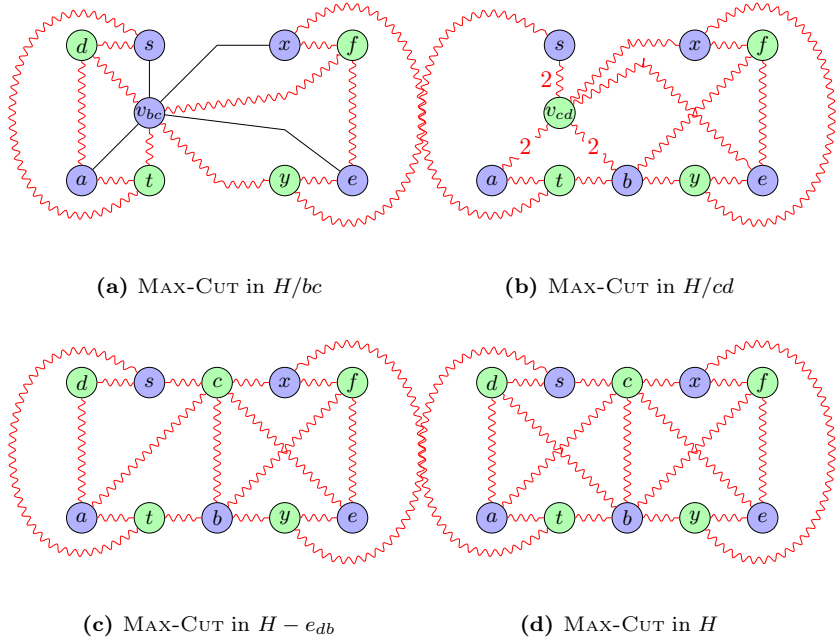
\begin{example}
	\label{bsp:bsp5}
	Let $H$ be the 2-almost-planar graph shown in Figure \ref{fig:rm_cr_0} with 
	uniform edge weights, e.g. 1. First the algorithm removes the crossing on the 
	left in the three described ways. The resulting graphs are shown in Figure 
	\ref{fig:rm_cr_4}-\ref{fig:rm_cr_2}. 
	For them the algorithm is called recursively. The first graph $H/bc$ is already 
	planar because the crossing on the right was dissolved when merging $b$ and $c$. 
	Therefore a planar \MCA{} is called. 
	For the other two graphs the algorithm creates six planar graphs, three each, 
	for which again a planar \MCA{} is called. The largest of the three calculated 
	cuts is then transfered back to the original graph $H/cd$ resp. $H-e_{bd}$.
	The recursively calculated cuts of the three graphs $H/bc$, $H/cd$ and $H-e_{bd}$ 
	are depicted in Figure \ref{fig:kfplMC_4}-\ref{fig:kfplMC_2} with \ref{fig:kfplMC_3} 
	showing the largest cut. This cut is transferred back to $H$ by splitting the 
	contracted node $v_{cd}$. The resulting cut is shown in Figure \ref{fig:kfplMC_MC}. 
	It is a maximum cut in $H$.
\end{example}

\subsection{Analysis}
\label{suse:alg-1pl:ana}

\begin{figure}[tb]
	\centering
	\begin{subfigure}{.115\linewidth}
		\centering
		\scalebox{0.55}{%
\begin{tikzpicture}
	\node[draw,circle,minimum size=0.6cm, inner sep=0ex, nodeset one] (a) at (0,0) {\Large $v$};
	\node[draw,circle,minimum size=0.6cm, inner sep=0ex, nodeset one] (b) at (2,0) {\Large $w$};
	\node[draw,circle,minimum size=0.6cm, inner sep=0ex, nodeset one] (c) at (2,2) {\Large $y$};
	\node[draw,circle,minimum size=0.6cm, inner sep=0ex, nodeset one] (d) at (0,2) {\Large $z$};
	
	\draw (a) -- (b);
	\draw (a) -- (c);
	\draw (a) -- (d);
	\draw (b) -- (c);
	\draw (b) -- (d);
	\draw (c) -- (d);
\end{tikzpicture} %
}
		\caption{\label{fig:8sepOption0}}
	\end{subfigure}
	\begin{subfigure}{.115\linewidth}
		\centering		
		\scalebox{0.55}{%
\begin{tikzpicture}
	\node[draw,circle,minimum size=0.5cm, inner sep=0ex, nodeset one] (a) at (0,0) {};
	\node[draw,circle,minimum size=0.5cm, inner sep=0ex, nodeset two] (b) at (2,0) {};
	\node[draw,circle,minimum size=0.5cm, inner sep=0ex, nodeset two] (c) at (2,2) {};
	\node[draw,circle,minimum size=0.5cm, inner sep=0ex, nodeset two] (d) at (0,2) {};
	
	\draw[red,line width=2pt,cut edge] (a) -- (b);
	\draw[red,line width=2pt,cut edge] (a) -- (c);
	\draw[red,line width=2pt,cut edge] (a) -- (d);
	\draw (b) -- (c);
	\draw (b) -- (d);
	\draw (c) -- (d);
\end{tikzpicture} %
}
		\caption{\label{fig:8sepOption1a}}
	\end{subfigure}
	\begin{subfigure}{.115\linewidth}
		\centering		
		\scalebox{0.55}{%
\begin{tikzpicture}
	\node[draw,circle,minimum size=0.5cm, inner sep=0ex, nodeset two] (a) at (0,0) {};
	\node[draw,circle,minimum size=0.5cm, inner sep=0ex, nodeset one] (b) at (2,0) {};
	\node[draw,circle,minimum size=0.5cm, inner sep=0ex, nodeset two] (c) at (2,2) {};
	\node[draw,circle,minimum size=0.5cm, inner sep=0ex, nodeset two] (d) at (0,2) {};
	
	\draw[red,line width=2pt,cut edge] (a) -- (b);
	\draw (a) -- (c);
	\draw (a) -- (d);
	\draw[red,line width=2pt,cut edge] (b) -- (c);
	\draw[red,line width=2pt,cut edge] (b) -- (d);
	\draw (c) -- (d);
\end{tikzpicture} %
}
		\caption{\label{fig:8sepOption1b}}
	\end{subfigure}	
	\begin{subfigure}{.115\linewidth}
		\centering
		\scalebox{0.55}{%
\begin{tikzpicture}
	\node[draw,circle,minimum size=0.5cm, inner sep=0ex, nodeset two] (a) at (0,0) {};
	\node[draw,circle,minimum size=0.5cm, inner sep=0ex, nodeset two] (b) at (2,0) {};
	\node[draw,circle,minimum size=0.5cm, inner sep=0ex, nodeset one] (c) at (2,2) {};
	\node[draw,circle,minimum size=0.5cm, inner sep=0ex, nodeset two] (d) at (0,2) {};

	\draw (a) -- (b);
	\draw[red,line width=2pt,cut edge] (a) -- (c);
	\draw (a) -- (d);
	\draw[red,line width=2pt,cut edge] (b) -- (c);
	\draw (b) -- (d);
	\draw[red,line width=2pt,cut edge] (c) -- (d);
\end{tikzpicture} %
}
		\caption{\label{fig:8sepOption1c}}
	\end{subfigure}
	\begin{subfigure}{.115\linewidth}
		\centering		
		\scalebox{0.55}{%
\begin{tikzpicture}
	\node[draw,circle,minimum size=0.5cm, inner sep=0ex, nodeset two] (a) at (0,0) {};
	\node[draw,circle,minimum size=0.5cm, inner sep=0ex, nodeset two] (b) at (2,0) {};
	\node[draw,circle,minimum size=0.5cm, inner sep=0ex, nodeset two] (c) at (2,2) {};
	\node[draw,circle,minimum size=0.5cm, inner sep=0ex, nodeset one] (d) at (0,2) {};

	\draw (a) -- (b);
	\draw (a) -- (c);
	\draw[red,line width=2pt,cut edge] (a) -- (d);
	\draw (b) -- (c);
	\draw[red,line width=2pt,cut edge] (b) -- (d);
	\draw[red,line width=2pt,cut edge] (c) -- (d);
\end{tikzpicture} %
}
		\caption{\label{fig:8sepOption1d}}
	\end{subfigure}
	\begin{subfigure}{.115\linewidth}
		\centering
		\scalebox{0.55}{%
\begin{tikzpicture}
	\node[draw,circle,minimum size=0.5cm, inner sep=0ex, nodeset one] (a) at (0,0) {};
	\node[draw,circle,minimum size=0.5cm, inner sep=0ex, nodeset one] (b) at (2,0) {};
	\node[draw,circle,minimum size=0.5cm, inner sep=0ex, nodeset two] (c) at (2,2) {};
	\node[draw,circle,minimum size=0.5cm, inner sep=0ex, nodeset two] (d) at (0,2) {};

	\draw (a) -- (b);
	\draw[red,line width=2pt,cut edge] (a) -- (c);
	\draw[red,line width=2pt,cut edge] (a) -- (d);
	\draw[red,line width=2pt,cut edge] (b) -- (c);
	\draw[red,line width=2pt,cut edge] (b) -- (d);
	\draw (c) -- (d);
\end{tikzpicture} %
}
		\caption{\label{fig:8sepOption2h}}
	\end{subfigure}
	\begin{subfigure}{.115\linewidth}
		\centering		
		\scalebox{0.55}{%
\begin{tikzpicture}
	\node[draw,circle,minimum size=0.5cm, inner sep=0ex, nodeset two] (a) at (0,0) {};
	\node[draw,circle,minimum size=0.5cm, inner sep=0ex, nodeset one] (b) at (2,0) {};
	\node[draw,circle,minimum size=0.5cm, inner sep=0ex, nodeset one] (c) at (2,2) {};
	\node[draw,circle,minimum size=0.5cm, inner sep=0ex, nodeset two] (d) at (0,2) {};

	\draw[red,line width=2pt,cut edge] (a) -- (b);
	\draw[red,line width=2pt,cut edge] (a) -- (c);
	\draw (a) -- (d);
	\draw (b) -- (c);
	\draw[red,line width=2pt,cut edge] (b) -- (d);
	\draw[red,line width=2pt,cut edge] (c) -- (d);
\end{tikzpicture} %
} 
		\caption{\label{fig:8sepOption2v}}
	\end{subfigure}
	\begin{subfigure}{.115\linewidth}
		\centering		
		\scalebox{0.55}{%
\begin{tikzpicture}
	\node[draw,circle,minimum size=0.5cm, inner sep=0ex, nodeset one] (a) at (0,0) {};
	\node[draw,circle,minimum size=0.5cm, inner sep=0ex, nodeset two] (b) at (2,0) {};
	\node[draw,circle,minimum size=0.5cm, inner sep=0ex, nodeset one] (c) at (2,2) {};
	\node[draw,circle,minimum size=0.5cm, inner sep=0ex, nodeset two] (d) at (0,2) {};

	\draw[red,line width=2pt,cut edge] (a) -- (b);
	\draw (a) -- (c);
	\draw[red,line width=2pt,cut edge] (a) -- (d);
	\draw[red,line width=2pt,cut edge] (b) -- (c);
	\draw (b) -- (d);
	\draw[red,line width=2pt,cut edge] (c) -- (d);
\end{tikzpicture} %
}
		\caption{\label{fig:8sepOption2d}}
	\end{subfigure}
	\caption{
		An illustration of the 8 non-isomorphic partitions of the four endpoints of a crossing \dahn{illustrated on $K_4$.
		Note that in general, the node pairs $vw, wy, yz$ and $zv$ might be connected by an edge, an arbitrary path or not at all.}
		(The red and curvy edges belong to the cut that is defined by the corresponding partition on $K_4$.)
	}
	\label{fig:8sepOptions}
\end{figure}
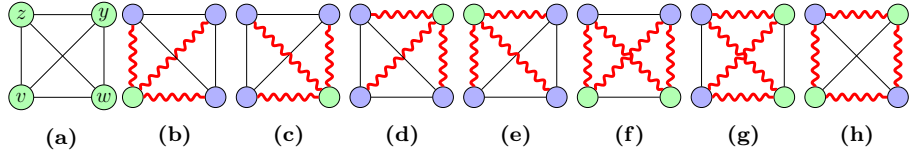

The four endpoints of a crossing can be partitioned in eight non-isomorphic ways,
cf.\@ Figure \ref{fig:8sepOptions}: 
(a) all endpoints in one set,
(b)/(c)/(d)/(e) three endpoints in one set without $v$/$w$/$y$/$z$,
(f)/(g) two endpoints of different crossing edges in one set each, or
(h) the two endpoints of the same crossing edge in one set each. 
For arbitrary graphs the induced cut might differ from the cuts shown in Figure \ref{fig:8sepOptions} because non-crossing edges might be replaced with a path or might not exist at all.

\begin{lemma}
	\label{lem:cut-transfer-from-G}	
	Let $G=(V,E,c)$ be a 1-planar graph with a 1-planar embedding $(\I,\Pi)$, $S\subseteq V$, and $\kz = \{e_{vy}, e_{wz}\} \in \I$ be an arbitrary crossing.
	\begin{enumerate}[(i)]
		\item Let $\Cut$ separate the four endpoints of $\kz$ in $G$ as shown in Figure \ref{fig:8sepOptions} (a), (b), (e) or (g). Then $\nodeCutExt{_1} = \nodeCutExt{ / wy}$ defines a cut in $G/wy$ with the same value. If $\Cut$ is maximum in $G$, so is $\CutExt{_1, G/wy}$ in $G/wy$.
		\label{lem:cut-transfer-from-G_contraction_wy}
		\item Let $\Cut$ separate the four endpoints of $\kz$ in $G$ as shown in Figure \ref{fig:8sepOptions} (a), (b), (c) or (f). Then $\nodeCutExt{_2} = \nodeCutExt{ / yz}$ defines a cut in $G/yz$ with the same value.  If $\Cut$ is maximum in $G$, so is $\CutExt{_2, G/yz}$ in $G/yz$.
		\label{lem:cut-transfer-from-G_contraction_yz}
		\item Let $\Cut$ separate the four endpoints of $\kz$ in $G$ as shown in Figure \ref{fig:8sepOptions} (a), (b), (d) or (h). Then $\nodeCutExt{_3} = \nodeCut$ defines a cut in $G - e_{wz}$ with the same value. If $\Cut$ is maximum in $G$, so is $\CutExt{_3, G - e_{wz}}$ in $G - e_{wz}$.
		\label{lem:cut-transfer-from-G_removal_wz}
	\end{enumerate}
\end{lemma}
\begin{proof}
	\textit{(\ref{lem:cut-transfer-from-G_contraction_wy})} Let $\nodeCut$ define a cut in $G$ that separates the endpoints of $\kz$ as shown in Figure \ref{fig:8sepOptions} (a), (b), (e) or (g).
	By contracting $w$ and $y$, the set of nodes is projected to $G/wy$ and $\CutExt{_1, G/wy}$ is a cut in $G/wy$. 
	The only edge that might have been removed in $G/wy$ does not add to the value of $\Cut$ in $G$ because $w$ and $y$ are not separated by the cut (cf. Fig. \ref{fig:8sepOptions} (a), (b), (e) or (g)). %
	Therefore the two cuts have the same value in both graphs. %
	Let $\nodeCut$ define a maximum cut in $G$ (with the required property). If there was a cut $\CutExt{',G/wy}$ in $G/wy$ larger than $\CutExt{_1, G/wy}$, then $\Split(\nodeCutExt{'},v_{wy})$ would define a cut in $G$ with the same value as $\CutExt{',G}$ (Lemma \ref{lem:cut-transfer-to-G} \ref{lem:cut-transfer-to-G_contraction}), contradicting that $\Cut$ is maximum in $G$.\\
	\textit{(\ref{lem:cut-transfer-from-G_contraction_yz})} The proof of the second proposition is analogous to the proof of the first.\\
	\textit{(\ref{lem:cut-transfer-from-G_removal_wz})} Let $\nodeCut$ define a cut in $G$ that separates the endpoints of $\kz$ as shown in Figure \ref{fig:8sepOptions} (a), (b), (d) or (h).
	Since $G$ and $G - e_{wz}$ have the same set of nodes, $\CutExt{_3,G - e_{wz}}$ is a cut in $G - e_{wz}$ as well.
	We know that $w$ and $z$ are not separated by the cut (cf. Fig. \ref{fig:8sepOptions} (a), (b), (d) or (h)). %
	Therefore the only edge that was removed in $G - e_{wz}$ does not add to the value of the cut in $G$ and the cut has the same value in both graphs.
	Let $\nodeCut$ define a maximum cut in $G$ (with the required property). If there was a cut $\CutExt{',G - e_{wz}}$ in $G - e_{wz}$ larger than $\CutExt{_3,G - e_{wz}}$, then $\CutExt{',G}$ would be a cut in $G$ as well (Lemma \ref{lem:cut-transfer-to-G} \ref{lem:cut-transfer-to-G_removal}), contradicting that $\Cut$ is maximum in $G$.
\end{proof}

\begin{theorem}
	\label{theo:opt}
	Algorithm \ref{algo:MaxCut1pl} computes a maximum cut in a weighted 1-planar graph $G$ with non-negative edge weights, given a set of crossing edges $X$ in a 1-planar embedding of $G$.
\end{theorem}
\begin{proof}
	We prove its optimality by induction over  $k=\vert X \vert$. Let $G$ be a weighted $k$-almost-planar graph.
	For $k=0$ the given graph is planar. Thus a \MCA for planar graphs calculates a node set defining a maximum cut in $G$.
	For $k>0$ we show that \dahn{the value of} the calculated cut is not smaller than \dahn{the value of} a maximum cut in $G$.
	Let $\nodeCutExt{^*}$ define a maximum cut in $G$ and let $\Cut$ be the cut defined by the calculated node set $\nodeCut$.
	Let $G_1 = G/wy, G_2 = G/yz$ and $G_3 = G - e_{wz}$ be the $(k-1)$-almost-planar graphs (Lemma \ref{lem:k-1-al-pl}), whose cuts $\CutExt{_1, G_1}, \CutExt{_2, G_2}$ and $\CutExt{_3, G_3}$ were calculated recursively by the algorithm. 
	There are 8 possible ways for $\nodeCutExt{^*}$ to separate the four endpoints of $\kz$. These are shown in Figure \ref{fig:8sepOptions} (a)--(h).
	If the endpoints of $\kz$ are separated as shown in (a), (b), (e) or (g), $\CutExt{^*, G}$ has the same value as a maximum cut $\CutExt{_1^*, G_1}$ in $G_1$ (Lemma \ref{lem:cut-transfer-from-G} \ref{lem:cut-transfer-from-G_contraction_wy}). Due to the induction hypothesis (IH), $\dahn{\wert(\CutExt{_1, G_1})}$ is not smaller than $\dahn{\wert(\CutExt{_1^*, G_1})}$. 
	If the endpoints of $\kz$ are separated as shown in (c) or (f), $\CutExt{^*, G}$ has the same value as a maximum cut $\CutExt{_2^*, G_2}$ in $G_2$ (Lemma \ref{lem:cut-transfer-from-G} \ref{lem:cut-transfer-from-G_contraction_yz}). Due to the induction hypothesis, $\dahn{\wert(\CutExt{_2, G_2})}$ is not smaller than $\dahn{\wert(\CutExt{_2^*, G_2})}$. 
	If the endpoints of $\kz$ are separated as shown in (d) or (h), $\CutExt{^*, G}$ has the same value as a maximum cut $\CutExt{_3^*, G_3}$ in $G_3$ (Lemma \ref{lem:cut-transfer-from-G} \ref{lem:cut-transfer-from-G_removal_wz}). Due to the induction hypothesis, $\dahn{\wert(\CutExt{_3, G_3})}$ is not smaller than $\dahn{\wert(\CutExt{_3^*, G_3})}$. 
	The algorithm chooses the node set defining the \dahn{cut with the highest value} (line \ref{algo:MaxCut1pl-old:maxGew}--\ref{algo:MaxCut1pl-old:convert_E}) 
	and transfers it back to $G$ without changing its value (Lemma \ref{lem:cut-transfer-to-G}). Thus \dahn{the value of} the calculated cut $\Cut$ is not smaller than $\dahn{\wert(\CutExt{^*, G})}$ and \dahn{$\Cut$} is therefore maximum in $G$. %
\end{proof}

The following equation summarizes the argumentation of the previous proof in the corresponding case, with $i = 1,2,3$:
$$\wert(\CutExt{^*, G}) 
\stackrel{L\ref{lem:cut-transfer-from-G}}{=} \wert(\CutExt{_i^*,G_i}) 
\stackrel{IH}{\leq} \wert(\CutExt{_i,G_i}) 
\stackrel{\stackrel{A\ref{algo:MaxCut1pl}}{L\ref{lem:cut-transfer-to-G}}}{\leq} 
\wert(\Cut)$$
In order to apply Lemma \ref{lem:cut-transfer-to-G} (\ref{lem:cut-transfer-to-G_removal}) 
to $\CutExt{_3, G_3}$ we need to assure that the two nodes of the removed edge $e_{wz}$ 
are on the same side of the cut. In the proof the assumption can be made because the 
endpoints of $\kz$ are separated as shown in Figure \ref{fig:8sepOptions} (d) or (h).
For the algorithm, however, this means that it can only choose $\nodeCut_3$ if \dahn{the value of} the other two cuts are truly smaller \dahn{than the value of the cut defined by $\nodeCut_3$}.

\subsection{Running time}
\label{suse:alg-1pl:time}

Let $n$ be the number of nodes and $m$ be the number of edges of a given graph. 
It is well known that a 1-planar graph has at most $4n-8$ edges \cite{PachT97}. 
For an arbitrary 1-planar embedding, the number of crossings is bounded by $\frac{m}{2}$, 
since every edge can be crossed at most once and every crossing needs two edges. 
With the previous observation, we can establish a bound depending on the number of nodes: $k \leq 2n-4$. 

\begin{theorem}
	\label{theo:time_calc_mc_kap}
	Algorithm \ref{algo:MaxCut1pl} computes a maximum cut in an embedded non-negatively edge-weighted 1-~planar graph with $n$ nodes and $k$ crossings in time $\mathcal{O}(3^k \cdot n^{3/2}  \log n)$ if one of the planar \MCA suggested in \cite{LiersP12} or \cite{ShihWuKuo90} is used.
\end{theorem}
\begin{proof}
	Let $T(k,n)$ be the running time of Algorithm \ref{algo:MaxCut1pl} on an embedded non-negatively edge-weighted 1-planar graph $G$ with $n$ nodes, $m$ edges and $k$ crossings in the given embedding. 
	If $G$ is planar, our algorithm uses a planar \MCA{} with running time $T_{p}(n)$, resulting in $T(0,n) = T_{p}(n)$.
	$\contract$ has a linear running time of $\mathcal{O}(k)$, since every crossing in $\I$ needs to be checked only once.
	The contractions of $G/wy$ and $G/yz$ take time $\mathcal{O}(n + m)$ and the edge removal $G - e_{wz}$ takes time $\mathcal{O}(m)$. Reversing a contraction on a set of nodes $S_i$ with $\Split$ takes $\vert S_i\vert$ steps, resulting in a running time of $\mathcal{O}(n)$.
	Since $m$ is bound by $4n-8$ \cite{PachT97} and $k$ is bound by $2n-4$ (see above Theorem \ref{theo:time_calc_mc_kap}) the recursive running time is: 
	$T(k,n) = 3 \cdot T(k-1,n) + \mathcal{O}(k + n + m) = 3 \cdot T(k-1,n) + \mathcal{O}(n)$.
	An induction proof confirms the closed form of the running time is: $$T(k,n) = 3^k \cdot \left[T(0,n) + \sum_{i=1}^{k} 3^{-i} \cdot \mathcal{O}(n)\right]$$
	Since the geometric sum equals a value between 0 and 1, the overall running time is $\mathcal{O}(3^k \cdot (T_{p}(n) + n))$.
	Liers and Pardella \cite{LiersP12} as well as Shih et al.~\cite{ShihWuKuo90} describe a planar \MCA with a running time of $\mathcal{O}(n^{3/2} \cdot \log n)$, resulting in a concrete  running time of $\mathcal{O}(3^k \cdot n^{3/2}  \log n)$ for our algorithm.
\end{proof}

If the number of crossings $k$ in a 1-planar embedding is fixed, the running time of Algorithm \ref{algo:MaxCut1pl} is polynomial. However, in an arbitrary 1-planar embedding, $k$ is not fixed and the factor $3^k$ leads to an exponential worst case running time. But we can show that our algorithm is \textit{fixed-parameter tractable} with parameter $k$.
Since the running time of our algorithm can be split in an exponential part, depending only on the parameter $k$, ($3^k$) and a polynomial part in the size of the input graph ($T_{p}(n) + n$), the algorithm is \textit{fixed-parameter tractable} with parameter $k$.

\begin{corollary}
	\label{cor:fpt}
	The weighted \MCP{} on embedded 1-planar graphs with non-negative edge weights and with $k$ crossings in the given embedding is \textit{fixed-parameter tractable} with parameter $k$.
\end{corollary}

\section{Counter example for crossing edges with negative weights}
\label{se:co-ex-neg-w}

\begin{figure}[tb]
	\centering
	\begin{subfigure}{.32\linewidth}
		\centering
		\scalebox{.45}{%
\begin{tikzpicture}[example]
	\clipExample
	\begin{scope}[every node/.style=example node]
	    \node (A) [nodeset one] at (-2,2) {$a$};
	    \node (B) [nodeset two] at (0,4) {$b$};
	    \node (C) [nodeset two] at (0,0) {$c$};
	    \node (D) [nodeset one] at (4,4) {$d$};
	    \node (E) [nodeset two] at (4,0) {$e$};
	\end{scope}
	
	\begin{scope}[%
	              every node/.style=example edge node,
	              every edge/.style=example edge]
	              
	    \path [-] (A) edge[cut edge] node {\edgeAB} (B);
	    \path [-] (A) edge[cut edge] node {\edgeAC} (C);
	    \path [-] (A) edge[out=90,in=130] node {\edgeAD} (D);
	    \path [-] (A) edge[cut edge,out=-90,in=-130] node {\edgeAE} (E);
	    \path [-] (B) edge node {\edgeBC} (C);
	    \path [-] (B) edge[cut edge] node {\edgeBD} (D);
	    \path [-] (B) edge[bend right=20] node[crossing edge node] {\edgeBE} (E);
	    \path [-] (D) edge[bend left=20,cut edge] node[crossing edge node] {\edgeCD} (C);
	    \path [-] (C) edge node {\edgeCE} (E);
	    \path [-] (D) edge[cut edge] node {\edgeDE} (E);
	     
	\end{scope}
	\node [example node,nodeset two] at (0,0) {$c$};
\end{tikzpicture} %
}
		\caption{\MC{} $\delta(S_{31},H)$ in $H$ with value~\sumOriginalCutBCE}
		\label{sfig:co-ex-K_5}
	\end{subfigure}
	\hfill
	\begin{subfigure}{.32\linewidth}
		\centering
		\scalebox{.45}{%
\begin{tikzpicture}[example]
	\clipExample
	\begin{scope}[every node/.style=example node]
	    \node (A)  [nodeset one] at (-2,2) {$a$};
	    \node (B)  [nodeset one] at (0,4) {$b$};
	    \node (C)  [nodeset two] at (0,0) {$c$};
	    \node (DE) [nodeset one] at (2,2) {$v_{de}$};
	    
	\end{scope}
	
	\begin{scope}[%
	              every node/.style=example edge node,
	              every edge/.style=example edge]
	    \path [-] (A) edge  node {\edgeAB} (B);
	    \path [-] (A) edge[cut edge] node {\edgeAC} (C);
	    \path [-] (B) edge[cut edge] node {\edgeBC} (C);
	    \path [-] (B) edge  node {\edgeBDE} (DE);
	    \path [-] (C) edge[cut edge] node {\edgeCDE} (DE);
	    \path [-,example edge] (A) to [out=90,in=-180] ($(B)+(0,1)$) node {\edgeADE} to [out=0, in=90 ]  (DE);

	\end{scope}
	
	\begin{scope}[every node/.style={draw=none,circle},
			      every edge/.style={draw=none,very thick}]
	    \node (Invis) at (4,0){};
	    \path (A) edge [out=-90,in=-130] node {\phantom{\edgeAE}} (Invis);
	    \node (D) at (4,4) {};
	    \path [-] (A) edge[out=90,in=130] node {\phantom{\edgeAD}} (D);
	\end{scope}
\end{tikzpicture} %
}
		\caption{\MC{} $\delta(S_1,H/ de)$ in $H/de$ with value~\sumMergedDE}
		\label{sfig:co-ex-K_5-merged-de}
	\end{subfigure}
	\hfill
	\begin{subfigure}{.32\linewidth}
		\centering
		\scalebox{.45}{%
\begin{tikzpicture}[example]
	\clipExample
	\begin{scope}[every node/.style=example node]
	    \node (A)  [nodeset one] at (-2,2) {$a$};
	    \node (BD) [nodeset one] at (2,2) {$v_{bd}$};
	    \node (C)  [nodeset two] at (0,0) {$c$};
	    \node (E)  [nodeset one] at (4,0) {$e$};
	\end{scope}
	
	\begin{scope}[
	              every node/.style=example edge node,
	              every edge/.style=example edge]
	    \path [-] (A) edge  node {\edgeABD} (BD);
	    \path [-] (A) edge[cut edge] node {\edgeAC} (C);
	    \path [-] (A) edge[out=-90,in=-130] node {\edgeAE} (E);
	    \path [-] (BD) edge[cut edge] node {\edgeBDC} (C);
	    \path [-] (C) edge[cut edge] node {\edgeCE} (E);
	    \path [-] (BD) edge  node {\edgeBDE} (E);		
	\end{scope}
	
	\begin{scope}[every node/.style={draw=none,circle},
			      every edge/.style={draw=none,very thick}]
		\draw [draw=none](A) to [out=90,in=-180] (0,5) node {\phantom{\edgeADE}} to [out=0, in=90 ]  (BD);
		\node (D) at (4,4) {};
	    \path [-] (A) edge[out=90,in=130] node {\phantom{\edgeAD}} (D);
		
	\end{scope}
\end{tikzpicture}

}
		\caption{\MC{} $\delta(S_1,H/bd)$ in $H/bd$ with value~\sumMergedBD}
		\label{sfig:co-ex-K_5-merged-bd}
	\end{subfigure}
	\begin{subfigure}{.32\linewidth}
		\centering
		\scalebox{.45}{%
\begin{tikzpicture}[example]
	\clipExample
	\begin{scope}[every node/.style=example node]
	    \node (A) [nodeset one] at (-2,2) {$a$};
	    \node (B) [nodeset two] at (0,4) {$b$};
	    \node (C) [nodeset two] at (0,0) {$c$};
	    \node (D) [nodeset one] at (4,4) {$d$};
	    \node (E) [nodeset two] at (4,0) {$e$};
	\end{scope}
	
	\begin{scope}[%
	              every node/.style=example edge node,
	              every edge/.style=example edge]
	    \path [-] (A) edge[cut edge] node {\edgeAB} (B);
	    \path [-] (A) edge[cut edge] node {\edgeAC} (C);
	    \path [-] (A) edge[out=90,in=130] node {\edgeAD} (D);
	    \path [-] (A) edge[cut edge,out=-90,in=-130] node {\edgeAE} (E);
	    \path [-] (B) edge  node {\edgeBC} (C);
	    \path [-] (B) edge[cut edge] node {\edgeBD} (D);
	    \path [-] (C) edge[cut edge] node {\edgeCD} (D);
	    \path [-] (C) edge  node {\edgeCE} (E);
	    \path [-] (D) edge[cut edge] node {\edgeDE} (E);
	     
	\end{scope}
\end{tikzpicture} %
}
		\caption{cut $\delta(S_{31},H-e_{be})$ with\\ value~\sumRemovedEdgeOne}
		\label{sfig:co-ex-K_5-cd-1}
	\end{subfigure}
	\hfill
	\begin{subfigure}{.32\linewidth}
		\centering
		\scalebox{.45}{%
\begin{tikzpicture}[example]
	\clipExample
	\begin{scope}[every node/.style=example node]
	    \node (A) [nodeset one] at (-2,2) {$a$};
	    \node (B) [nodeset two] at (0,4) {$b$};
	    \node (C) [nodeset one] at (0,0) {$c$};
	    \node (D) [nodeset one] at (4,4) {$d$};
	    \node (E) [nodeset two] at (4,0) {$e$};
	\end{scope}
	
	\begin{scope}[%
	              every node/.style=example edge node,
	              every edge/.style=example edge]
	    \path [-] (A) edge[cut edge] node {\edgeAB} (B);
	    \path [-] (A) edge  node {\edgeAC} (C);
	    \path [-] (A) edge[out=90,in=130] node {\edgeAD} (D);
	    \path [-] (A) edge[cut edge,out=-90,in=-130] node {\edgeAE} (E);
	    \path [-] (B) edge[cut edge] node {\edgeBC} (C);
	    \path [-] (B) edge[cut edge] node {\edgeBD} (D);
	    \path [-] (C) edge  node {\edgeCD} (D);
	    \path [-] (C) edge[cut edge] node {\edgeCE} (E);
	    \path [-] (D) edge[cut edge] node {\edgeDE} (E);
	     
	\end{scope}
\end{tikzpicture} %
}
		\caption{cut $\delta(S_{32},H-e_{be})$ with\\ value~\sumRemovedEdgeTwo}
		\label{sfig:co-ex-K_5-cd-2}
	\end{subfigure}
	\hfill
	\begin{subfigure}{.32\linewidth}
		\centering
		\scalebox{.45}{%
\begin{tikzpicture}[example]
	\clipExample
	\begin{scope}[every node/.style=example node]
	    \node (A) [nodeset one] at (-2,2) {$a$};
	    \node (B) [nodeset two] at (0,4) {$b$};
	    \node (C) [nodeset two] at (0,0) {$c$};
	    \node (D) [nodeset one] at (4,4) {$d$};
	    \node (E) [nodeset one] at (4,0) {$e$};
	\end{scope}
	
	\begin{scope}[%
	              every node/.style=example edge node,
	              every edge/.style=example edge]
	    \path [-] (A) edge[cut edge] node {\edgeAB} (B);
	    \path [-] (A) edge[cut edge] node {\edgeAC} (C);
	    \path [-] (A) edge[out=90,in=130] node {\edgeAD} (D);
	    \path [-] (A) edge[out=-90,in=-130] node {\edgeAE} (E);
	    \path [-] (B) edge  node {\edgeBC} (C);
	    \path [-] (B) edge[cut edge] node {\edgeBD} (D);
	    \path [-] (C) edge[cut edge] node {\edgeCD} (D);
	    \path [-] (C) edge[cut edge] node {\edgeCE} (E);
	    \path [-] (D) edge  node {\edgeDE} (E);
	     
	\end{scope}
\end{tikzpicture} %
}
		\caption{cut $\delta(S_{33},H-e_{be})$ with\\ value~\sumRemovedEdgeThree\ (but value~\sumRemovedEdgeThreeTrueVal{} in $H$!)}
		\label{sfig:co-ex-K_5-cd-3}
	\end{subfigure}
	\caption{A counter example, showing that Algorithm \ref{algo:MaxCut1pl} might fail to calculate a maximum cut in a graph, which has a crossings with two negative edges.}
	\label{fig:counter_example_1}
\end{figure}
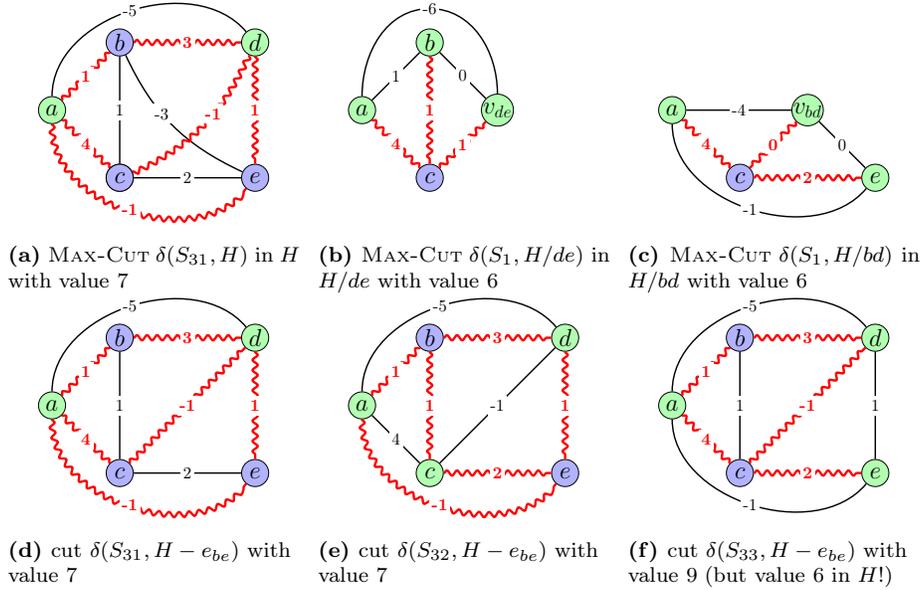

Considering graphs with arbitrary edge weights the algorithm introduced in Section \ref{se:alg-1pl} might not find the optimal solution. If a deleted edge had a negative weight 
it is possible that the planar \MCA chooses a maximum cut, which is not optimal in the original graph. 
This can happen if %
the calculated maximum cut separates the nodes of the removed edge. %
When transfered back to the original graph the negative weight of the before removed edge is added, decreasing the value of the cut.
Hence the optimality can not be guaranteed.
If the optimal cut can only be found in the subgraph $G_3$ and the cut separating the nodes of the removed edge is the only maximum cut in the planar subgraph, Algorithm \ref{algo:MaxCut1pl} will then never find the maximum cut in $G$.
An example is given in Figure \ref{fig:counter_example_1}. 

\begin{example}
	Let our graph $H$ be the embedded weighted $K_5$ as shown in Figure \ref{sfig:co-ex-K_5} with a maximum cut value of \sumOriginalCutBCE. One possible maximum cut in $H$ is defined by $S_{31}=\{b,c,e\}$ (as shown in Fig. \ref{sfig:co-ex-K_5}) another is defined by $S_{32}=\{b,e\}$.
	The 1-planar \MCA introduced in Section \ref{se:alg-1pl} is based on the observation that the largest of the maximum cuts in $H/de$, $H/bd$ and $H-e_{be}$  has a corresponding cut in $H$ of equal value, which is maximum in $H$ (cf. Lemma \ref{lem:cut-transfer-to-G} and Theorem \ref{theo:opt}). 
	It is easy to see that the maximum cuts in $H/de$ and $H/bd$ are not maximum in $H$ %
	(cf. Fig. \ref{sfig:co-ex-K_5-merged-de} and Fig. \ref{sfig:co-ex-K_5-merged-bd}). Therefore a maximum cut in $H$ has to be found in $H-e_{be}$.

	If we look at the two before mentioned cuts, defined by $S_{31}$ and $S_{32}$, in $H-e_{be}$, we can easily see that both have the same cut value in $H-e_{be}$ and in $H$, which is \sumOriginalCutBCE{} (cf. Fig. \ref{sfig:co-ex-K_5-cd-1} and Fig. \ref{sfig:co-ex-K_5-cd-2}).
	But the maximum cut in $H-e_{be}$ has a value of \sumRemovedEdgeThree{} and is defined by $S_{33}=\{b,c\}$ (cf. Fig. \ref{sfig:co-ex-K_5-cd-3}). 
	The partition $S_{33}$ separates $b$ and $e$, lowering the cut value in $H$ by $c(e_{be})=-3$ if after the cut is transfered back to $H$. Therefore its cut value in $H$ is only \sumRemovedEdgeThreeTrueVal{}.
	The algorithm would choose $S_{33}$ as the maximum cut of $H$ because it has the highest cut value of the three recursively calculated cuts.
	But in $H$ the cut value is only \sumRemovedEdgeThreeTrueVal{} due to the weight of the negative edge that was removed in $H-e_{be}$.

	One could attempt to solve this problem by calculating the value of a cut, found in the subgraphs, on the original graph.
	However, this still does not ensure that the algorithm chooses the correct cut.
	While this would prevent the algorithm from choosing the wrong cut $S_{33}$, it would choose $S_1=\{c\}$ instead, because it does not recognize there are other cuts in $H-e_{be}$ with a higher value in $H$. 
	Note that the maximum cut in $H$ can only be found in $H-e_{be}$ (compare Fig. \ref{sfig:co-ex-K_5} and \ref{sfig:co-ex-K_5-cd-1}). 
	But the algorithm cannot find it, because $e_{be}$ has a negative edge weight.%
\end{example}

Algorithm \ref{algo:MaxCut1pl} is based on the observation that the largest of the maximum cuts in $G/wy$, $G/yz$ and $G-e_{wz}$ has a corresponding cut in $G$ of equal value, which is maximum in $G$ (cf. Theorem \ref{theo:opt}). 
This only holds for 1-planar graphs with non-negative edge weights because for those graphs we can ensure that a planar \MCA calculates a maximum cut, which does not decrease in value when transferred back to $G$.
As long as only one of the crossing edges has a negative weight we could avoid this problem by deleting the crossing edge with the non-negative weight and choosing the to be contracted nodes accordingly. 
This results from the fact that, by reintroducing the deleted edge into our graph, the value of the cut can only increase.
If both edges of a crossing have negative edge weights, however, we need to utilize a more elaborated solution, which we introduce in the following section.

\section{Max-Cut for embedded 1-planar graphs with arbitrary edge weights}
\label{se:alg-neg}

\begin{figure}[tb]
	\centering
	\begin{subfigure}[b]{.32\linewidth}
		\centering
		\scalebox{1}{%
\begin{tikzpicture}
	\node[draw,circle,minimum size=0.5cm, inner sep=0ex] (a) at (0,0) {$v$};
	\node[draw,circle,minimum size=0.5cm, inner sep=0ex] (b) at (2,0) {$w$};
	\node[draw,circle,minimum size=0.5cm, inner sep=0ex] (c) at (2,2) {$y$};
	\node[draw,circle,minimum size=0.5cm, inner sep=0ex] (d) at (0,2) {$z$};

	\draw (a) -- (b);
	\draw (a) -- (c);
	\draw (a) -- (d);
	\draw (b) -- (c);
	\draw (b) -- (d);
	\draw (c) -- (d);
	
	\node[draw,circle,fill=white,minimum size=0.2cm, inner sep=0ex] at (1,2) {};
	\node[draw,circle,fill=white,minimum size=0.2cm, inner sep=0ex] at (1,0) {};
\end{tikzpicture} %
}
		\caption{A crossing.}
		\label{fig:G4_1}
	\end{subfigure}
	\begin{subfigure}[b]{.32\linewidth}
		\centering
		\scalebox{1}{%
\begin{tikzpicture}
	\node[draw,circle,circle,minimum size=0.5cm, inner sep=0ex] (a) at (0,0) {$v$};
	\node[draw,circle,minimum size=0.5cm, inner sep=0ex] (b) at (2,0) {$w$};
	\node[draw,circle,circle,minimum size=0.5cm, inner sep=0ex] (c) at (2,2) {$y$};
	\node[draw,circle,circle,minimum size=0.5cm, inner sep=0ex] (d) at (0,2) {$z$};

	\draw (a) -- (b);
	\draw (a) -- (c);
	\draw (a) -- (d);
	\draw (b) -- (c);
	\draw (c) -- (d);
	
	\node[draw,circle,fill=white,minimum size=0.2cm, inner sep=0ex] at (1,2) {};
	\node[draw,circle,fill=white,minimum size=0.2cm, inner sep=0ex] at (1,0) {};
\end{tikzpicture} %
}
		\caption{Crossing edge removed.}
		\label{fig:G4_2}
	\end{subfigure}
	\begin{subfigure}[b]{.32\linewidth}
		\centering
		\scalebox{1}{%
\begin{tikzpicture}
	\node[draw,circle,minimum size=0.5cm, inner sep=0ex] (a) at (0,0) {$v$};
	\node[draw,circle,minimum size=0.5cm, inner sep=0ex, nodeset two] (b) at (2,0) {$w$};
	\node[draw,circle,minimum size=0.5cm, inner sep=0ex, nodeset one] (c) at (2,2) {$y$};
	\node[draw,circle,minimum size=0.5cm, inner sep=0ex, nodeset two] (d) at (0,2) {$z$};

	\draw (a) -- (b);
	\draw (a) -- (c);
	\draw (a) -- (d);
	\draw (b) -- (c);
	\draw (c) -- (d);
	
	\draw [blue, thick]
	 (b) to (c);
	\draw [red, thick, %
	bend left=30](c) to (d);

	\node[draw,circle,fill=white,minimum size=0.2cm, inner sep=0ex] at (1,2) {};
	\node[draw,circle,fill=white,minimum size=0.2cm, inner sep=0ex] at (1,0) {};
\end{tikzpicture} %
}
		\caption{New edge $e_{yz}$ added.}
		\label{fig:G4_3}
	\end{subfigure}
	\caption{Removing a crossing and adding edges to ensure that $\{w,z\}$ and $y$ are on opposite sides of the cut (enforcing a cut as seen in Fig. \ref{fig:8sepOption1c} or \ref{fig:8sepOption2d}). The red edge $e_{yz}$ -- which was added to the graph -- and the blue edge $e_{wy}$ -- which was part of the original graph -- are forced to be in the calculated cut, ensuring the separation of $\{w,z\}$ and $y$.}
	\label{fig:G4}
\end{figure}
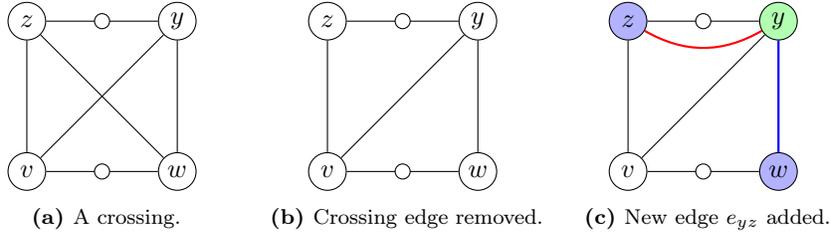

To deal with the problems arising from negative weights on crossing edges, we make use of the insights we gained in Section \ref{se:alg-1pl}.
Our goal is to construct a graph that forces its maximum cut to partition the nodes of the crossing as depicted in Figure \ref{fig:8sepOption1c} or \ref{fig:8sepOption2d}. 
Therefore we need to ensure that $\{w,z\}$ and $y$ are separated by the cut.
First we remove the crossing edge $e_{wz}$ because it will not contribute to the cut (cf. Fig. \ref{fig:G4_2}). 
Then we insert the edges $e_{wy}$ and $e_{yz}$ if they do not already exist (cf. Fig. \ref{fig:G4_3}).
The newly inserted edges shall each have weight $0$, so they do not add to the value of the cut.
Finally the two edges $e_{wy}$ and $e_{yz}$ are added to a set of \textit{fixed cut edges} $C$.
Using the \textsc{FixedCutEdges}-\MCA\ (\FCEMC) introduced by Liers and Pardella \cite{LiersP12}, the edges in $C$ are forced to be in the calculated maximum cut, ensuring the separation of $\{w,z\}$ and $y$ (cf. Fig. \ref{fig:G4_3}).

\subsection{The Algorithm}
To be able to deal with negative weights on crossing edges we adapt the algorithm presented in Section \ref{se:alg-1pl}. 
Note that the new algorithm (cf. Algorithm \ref{algo:MaxCut1pl-neg}) behaves similar to Algorithm \ref{algo:MaxCut1pl}. The main difference is the set of fixed cut edges $C$.
Algorithm \ref{algo:MaxCut1pl-neg} expects a weighted 1-planar graph $G=(V,E,c)$, a set of crossings $\I$, and a set of edges $C\subset E$ as input. 
The set $C$ contains the edges that are forced to be in the calculated cut. It is initialized with $C=\emptyset$.
\begin{algorithm}

\underline{$\kfplMCb(G,\I,C)$}\vspace{0.5em}\\
\algorithmicrequire \ %
An undirected weighted 1-planar graph $G=(V,E,c)$, a set of crossing edges $\I$ in a 1-planar embedding of $G$ and a set of edges the algorithm has to cut $C \subseteq E$.\\
\algorithmicensure \ A set $\nodeCut \subseteq V$ defining a maximum cut $\Cut \subseteq E$ in $G$.

\begin{algorithmic}[1]
\IF{$\I = \emptyset$}
	\STATE $\nodeCut \gets \FCEMC(G,C)$  		\COMMENT{\textsc{FixedCutEdges-Max-Cut} \cite{LiersP12}}			\label{algo:MaxCut1pl-neg:plMC}
	\ELSE
	\STATE choose an element $\kz\gets \{e_{vy}, e_{wz}\} \in \I$
	\IF{$wy \notin C$}																		\label{algo:MaxCut1pl-neg:check1}
	\STATE $G_1 \gets G/wy$ %
	\STATE $\nodeCutExt{_1} \gets \kfplMCb(G_1, \contract(\I,w,y), C)$						\label{algo:MaxCut1pl-neg:f1}
	\ELSE
	\STATE $\nodeCutExt{_1} \gets \emptyset$
	\ENDIF
	\IF{$yz \notin C$}																		\label{algo:MaxCut1pl-neg:check2}
	\STATE $G_2 \gets G/yz$
	\STATE $\nodeCutExt{_2} \gets \kfplMCb(G_2, \contract(\I,y,z), C)$						\label{algo:MaxCut1pl-neg:f2}
	\ELSE
	\STATE $\nodeCutExt{_2} \gets \emptyset$
	\ENDIF
	\STATE $G_3 \gets \AddEdges(G - e_{wz},\{e_{wy}, e_{yz}\})$  \label{algo:MaxCut1pl-neg:add_edges} \COMMENT{add edges with weight 0 if not in $G$}
	\STATE $C' \gets C \cup \{e_{wy}, e_{yz}\}$ \label{algo:MaxCut1pl-neg:add_edges_to_cut}\COMMENT{set of fixed cut edges}
	\STATE $\nodeCutExt{_3} \gets \kfplMCb(G_3, \I \setminus \{\kz\},C')$ 		\label{algo:MaxCut1pl-neg:enforced_maxCut} 
	\STATE $j \gets \argmax{1\leq i\leq 3} \wert(\CutExt{_i,G_i})$				\label{algo:MaxCut1pl-neg:maxGew}	
	\IF{$j = 1$}																\label{algo:MaxCut1pl-neg:convert_S}
	\STATE $\nodeCut \gets \Split(\nodeCutExt{_1},v_{wy})$						\label{algo:MaxCut1pl-neg:KS1}
	\ELSIF{$j = 2$}
	\STATE $\nodeCut \gets \Split(\nodeCutExt{_2},v_{yz})$						\label{algo:MaxCut1pl-neg:KS2}	
	\ELSIF{$j = 3$}
	\STATE $\nodeCut \gets \nodeCutExt{_3}$										\label{algo:MaxCut1pl-neg:KS3}
	\ENDIF																		\label{algo:MaxCut1pl-neg:convert_E}
\ENDIF
\STATE \textbf{return} $\nodeCut$												\label{algo:MaxCut1pl-neg:return}
\end{algorithmic}
\caption{Weighted \MCA for embedded 1-planar graphs with arbitrary edge weights.} 	\label{algo:MaxCut1pl-neg}
\end{algorithm}

The new algorithm initially proceeds similarly to the old one: 
If there are no crossings left, the algorithm computes a planar \MC{} (line \ref{algo:MaxCut1pl-neg:plMC}) and returns it.
The difference is that the planar \textsc{FixedCutEdges}-\MCA{} (\FCEMC) from Liers and Pardella is used \cite{LiersP12}. It receives a set of edges $C$, which will be forced to be in the calculated maximum cut.
If a cut respecting $C$ is impossible, e.g., because a subset of $C$ forms a circle of uneven size, \FCEMC{} will fail to calculate a cut. In this case $S=\emptyset$ -- a cut with value 0 -- is returned. 
Contracting two nodes after their connecting edge was added to $C$, would also lead to a contradiction. 
In this case $S=\emptyset$ is returned as well.

The first two cuts get calculated exactly as they were in Algorithm \ref{algo:MaxCut1pl}: by a recursive call of \kfplMCb{} on $G/wy$ and $G/yz$. %
The third graph, $G_3$, is created by removing $e_{wz}$ from $G$ and adding the edges $e_{wy}$ and $e_{yz}$, with a weight of 0, if they do not already exist in $G$. %
This method is called \AddEdges{} (line \ref{algo:MaxCut1pl-neg:add_edges}).
The edges $e_{wy}$ and $e_{yz}$ get then added to the set of fixed cut edges $C$ %
(line \ref{algo:MaxCut1pl-neg:add_edges_to_cut}).
Finally the algorithm recursively calls \kfplMCb{} on $G_3$, using the adapted fixed cut edges set $C'$, and saves the return value in $S_3$ (line \ref{algo:MaxCut1pl-neg:enforced_maxCut}).
As described above, this ensures that $\{w,z\}$ and $y$ are separated by the cut, meaning: $\{w,z\} \subseteq S_3$ and $y\in \overline{S_3}$ or vice versa.
Afterwards the algorithm once again behaves the same way as Algorithm \ref{algo:MaxCut1pl}. %
It determines which cut has the highest value (line \ref{algo:MaxCut1pl-neg:maxGew}), chooses the according node set (lines \ref{algo:MaxCut1pl-neg:KS1} -- %
\ref{algo:MaxCut1pl-neg:KS3}), and returns it (line \ref{algo:MaxCut1pl-neg:return}).

As noted before, it is possible that the returned cut is $S=\emptyset$ because this is either indeed the maximum cut %
or because all calls of \kfplMCb{} (lines \ref{algo:MaxCut1pl-neg:f1}, \ref{algo:MaxCut1pl-neg:f2}, \ref{algo:MaxCut1pl-neg:enforced_maxCut}) received a set of fixed cut edges, which contradicts itself, e.g., because a subset of $C$ forms a circle of uneven size. %
If $S=\emptyset$, which will always have a value of 0, is chosen in line \ref{algo:MaxCut1pl-neg:maxGew}, %
every other cut has to have a non-positive value, making $S=\emptyset$ indeed the correct choice.%

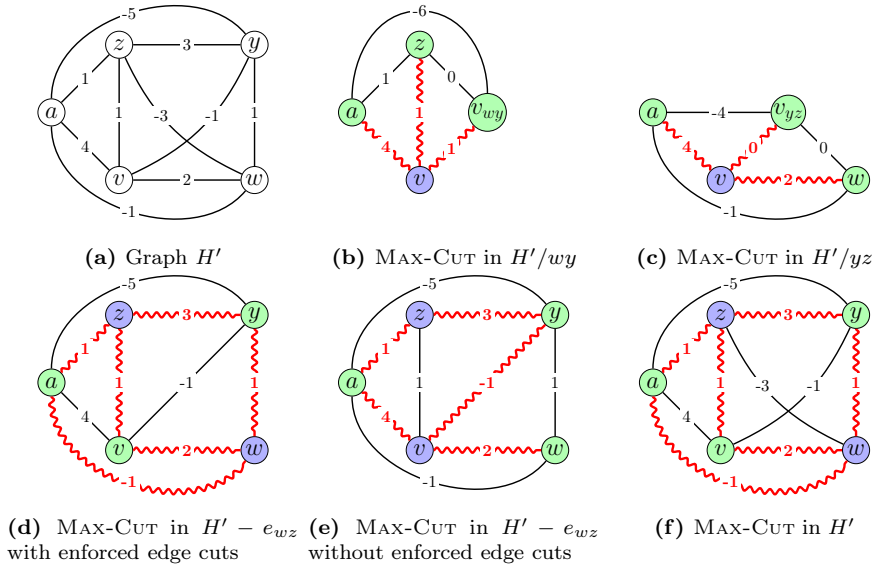
\begin{figure}[h!]
	\centering
	\begin{subfigure}{.32\linewidth}
		\centering
		\scalebox{.45}{%
\begin{tikzpicture}[example]
	\clipExample
	\begin{scope}[every node/.style=example node]
	    \node (A) at (-2,2) {$a$};
	    \node (B) at (0,4) {$z$};
	    \node (C) at (0,0) {$v$};
	    \node (D) at (4,4) {$y$};
	    \node (E) at (4,0) {$w$};
	\end{scope}
	
	\begin{scope}[every node/.style=example edge node,
	              every edge/.style=example edge]
	    \path [-] (A) edge[] node {\edgeAB} (B);
	    \path [-] (A) edge[] node {\edgeAC} (C);
	    \path [-] (A) edge[out=90,in=130] node {\edgeAD} (D);
	    \path [-] (A) edge[out=-90,in=-130] node {\edgeAE} (E);
	    \path [-] (B) edge node {\edgeBC} (C);
	    \path [-] (B) edge[] node {\edgeBD} (D);
	    \path [-] (B) edge[bend right=20] node[crossing edge node] {\edgeBE} (E);
	    \path [-] (D) edge[bend left=20] node[crossing edge node] {\edgeCD} (C);
	    \path [-] (C) edge node {\edgeCE} (E);
	    \path [-] (D) edge[] node {\edgeDE} (E);
	     
	\end{scope}
\end{tikzpicture} %
}
		\caption{Graph $H'$}
		\label{sfig:bsp_algo2_1}
	\end{subfigure}
	\begin{subfigure}{.32\linewidth}
		\centering
		\scalebox{.45}{%
\begin{tikzpicture}[example]
	\clipExample
	\begin{scope}[every node/.style=example node]
	    \node (A)  [nodeset one] at (-2,2) {$a$};
	    \node (B)  [nodeset one] at (0,4) {$z$};
	    \node (C)  [nodeset two] at (0,0) {$v$};
	    \node (DE) [nodeset one] at (2,2) {$v_{wy}$};
	    
	\end{scope}
	
	\begin{scope}[%
	              every node/.style=example edge node,
	              every edge/.style=example edge]
	    \path [-] (A) edge node {\edgeAB} (B);
	    \path [-] (A) edge[cut edge] node {\edgeAC} (C);
	    \path [-] (B) edge[cut edge] node {\edgeBC} (C);
	    \path [-] (B) edge node {\edgeBDE} (DE);
	    \path [-] (C) edge[cut edge] node {\edgeCDE} (DE);
	    \path [-,example edge] (A) to [out=90,in=-180] ($(B)+(0,1)$) node {\edgeADE} to [out=0, in=90 ]  (DE);

	\end{scope}
	
	\begin{scope}[every node/.style={draw=none,circle},
			      every edge/.style={draw=none,very thick}]
	    \node (Invis) at (4,0){};
	    \path (A) edge [out=-90,in=-130] node {\phantom{\edgeAE}} (Invis);
	    \node (D) at (4,4) {};
	    \path [-] (A) edge[out=90,in=130] node {\phantom{\edgeAD}} (D);
	\end{scope}
\end{tikzpicture} %
}
		\caption{\MC{} in $H'/wy$}
		\label{sfig:bsp_algo2_2}
	\end{subfigure}
	\begin{subfigure}{.32\linewidth}
		\centering
		\scalebox{.45}{%
\begin{tikzpicture}[example]
	\clipExample
	\begin{scope}[every node/.style=example node]
	    \node (A)  [nodeset one] at (-2,2) {$a$};
	    \node (BD) [nodeset one] at (2,2) {$v_{yz}$};
	    \node (C)  [nodeset two] at (0,0) {$v$};
	    \node (E)  [nodeset one] at (4,0) {$w$};
	\end{scope}
	
	\begin{scope}[
	              every node/.style=example edge node,
	              every edge/.style=example edge]
	    \path [-] (A) edge node {\edgeABD} (BD);
	    \path [-] (A) edge[cut edge] node {\edgeAC} (C);
	    \path [-] (A) edge[out=-90,in=-130] node {\edgeAE} (E);
	    \path [-] (BD) edge[cut edge] node {\edgeBDC} (C);
	    \path [-] (C) edge[cut edge] node {\edgeCE} (E);
	    \path [-] (BD) edge node {\edgeBDE} (E);		
	\end{scope}
	
	\begin{scope}[every node/.style={draw=none,circle},
			      every edge/.style={draw=none,very thick}]
		\draw [draw=none](A) to [out=90,in=-180] (0,5) node {\phantom{\edgeADE}} to [out=0, in=90 ]  (BD);
		\node (D) at (4,4) {};
	    \path [-] (A) edge[out=90,in=130] node {\phantom{\edgeAD}} (D);
		
	\end{scope}
\end{tikzpicture} %
}
		\caption{\MC{} in $H'/yz$}
		\label{sfig:bsp_algo2_3}
	\end{subfigure}
	\begin{subfigure}{.32\linewidth}
		\centering
		\scalebox{.45}{%
\begin{tikzpicture}[example]
	\clipExample
	\begin{scope}[every node/.style=example node]
	    \node (A) [nodeset one] at (-2,2) {$a$};
	    \node (B) [nodeset two] at (0,4) {$z$};
	    \node (C) [nodeset one] at (0,0) {$v$};
	    \node (D) [nodeset one] at (4,4) {$y$};
	    \node (E) [nodeset two] at (4,0) {$w$};
	\end{scope}
	
	\begin{scope}[%
	              every node/.style=example edge node,
	              every edge/.style=example edge]
	    \path [-] (A) edge[cut edge] node {\edgeAB} (B);
	    \path [-] (A) edge node {\edgeAC} (C);
	    \path [-] (A) edge[out=90,in=130] node {\edgeAD} (D);
	    \path [-] (A) edge[cut edge,out=-90,in=-130] node {\edgeAE} (E);
	    \path [-] (B) edge[cut edge] node {\edgeBC} (C);
	    \path [-] (B) edge[cut edge] node {\edgeBD} (D);
	    \path [-] (C) edge node {\edgeCD} (D);
	    \path [-] (C) edge[cut edge] node {\edgeCE} (E);
	    \path [-] (D) edge[cut edge] node {\edgeDE} (E);
	     
	\end{scope}
\end{tikzpicture} %
}
		\caption{\MC{} in $H'-e_{wz}$ with enforced edge cuts}
		\label{sfig:bsp_algo2_5}
	\end{subfigure}
	\begin{subfigure}{.32\linewidth}
		\centering
		\scalebox{.45}{%
\begin{tikzpicture}[example]
	\clipExample
	\begin{scope}[every node/.style=example node]
	    \node (A) [nodeset one] at (-2,2) {$a$};
	    \node (B) [nodeset two] at (0,4) {$z$};
	    \node (C) [nodeset two] at (0,0) {$v$};
	    \node (D) [nodeset one] at (4,4) {$y$};
	    \node (E) [nodeset one] at (4,0) {$w$};
	\end{scope}
	
	\begin{scope}[%
	              every node/.style=example edge node,
	              every edge/.style=example edge]
	    \path [-] (A) edge[cut edge] node {\edgeAB} (B);
	    \path [-] (A) edge[cut edge] node {\edgeAC} (C);
	    \path [-] (A) edge[out=90,in=130] node {\edgeAD} (D);
	    \path [-] (A) edge[out=-90,in=-130] node {\edgeAE} (E);
	    \path [-] (B) edge[] node {\edgeBC} (C);
	    \path [-] (B) edge[cut edge] node {\edgeBD} (D);
	    \path [-] (C) edge[cut edge] node {\edgeCD} (D);
	    \path [-] (C) edge[cut edge] node {\edgeCE} (E);
	    \path [-] (D) edge[] node {\edgeDE} (E);
	     
	\end{scope}
\end{tikzpicture} %
}
		\caption{\MC{} in $H'-e_{wz}$ without enforced edge cuts}
		\label{sfig:bsp_algo2_5-counter}
	\end{subfigure}
		\begin{subfigure}{.32\linewidth}
		\centering
		\scalebox{.45}{%
\begin{tikzpicture}[example]
	\clipExample
	\begin{scope}[every node/.style=example node]
	    \node (A) [nodeset one] at (-2,2) {$a$};
	    \node (B) [nodeset two] at (0,4) {$z$};
	    \node [nodeset one] (C) at (0,0) {$v$};
	    \node [nodeset one] (D) at (4,4) {$y$};
	    \node [nodeset two] (E) at (4,0) {$w$};
	\end{scope}
	
	\begin{scope}[%
	              every node/.style=example edge node,
	              every edge/.style=example edge]
	    \path [-] (A) edge [cut edge] node {\edgeAB} (B);
	    \path [-] (A) edge [] node {\edgeAC} (C);
	    \path [-] (A) edge [out=90,in=130] node {\edgeAD} (D);
	    \path [-] (A) edge [cut edge,out=-90,in=-130] node {\edgeAE} (E);
	    \path [-] (B) edge [cut edge] node {\edgeBC} (C);
	    \path [-] (B) edge [cut edge] node {\edgeBD} (D);
	    \path [-] (B) edge [bend right=20] node[crossing edge node] {\edgeBE} (E);
	    \path [-] (D) edge [bend left=20] node[crossing edge node] {\edgeCD} (C);
	    \path [-] (C) edge [cut edge] node {\edgeCE} (E);
	    \path [-] (D) edge [cut edge] node {\edgeDE} (E);
	     
	\end{scope}
\end{tikzpicture} %
}
		\caption{\MC{} in $H'$\\ \ }
		\label{sfig:bsp_algo2_6}
	\end{subfigure}
	\caption{An example how the algorithm calculates a \MC~in the embedded 1-almost-planar graph $H'$, containing a crossing with two negative weight edges.}
	\label{fig:bsp_algo2}
\end{figure}

\begin{example}
	\label{bsp:bsp_algo2}
	Let $H'$ be the 1-almost-planar graph shown in Figure \ref{sfig:bsp_algo2_1}.
	We obtain $H'$ by renaming the nodes of the graph $H$, used in our counter example in Section \ref{se:co-ex-neg-w} (cf. Figure \ref{sfig:co-ex-K_5}), according to how the algorithm chooses $w$, $y$ and $z$. 
	In this example we choose edge $e_{be}$ in $H$ as edge $e_{wz}$ and node $d$ in $H$ as node $y$. The last remaining node of the crossing is renamed from $c$ in $H$ to $v$ in $H'$.
	
	The  algorithm resolves the crossing in the three different ways explained above.
	The resulting graphs and their calculated maximum cuts are shown in Figures \ref{sfig:bsp_algo2_2} -- \ref{sfig:bsp_algo2_5}.
	The cuts shown in Figures \ref{sfig:bsp_algo2_2} and \ref{sfig:bsp_algo2_3} are calculated using the FCE-\MCA with $C=\emptyset$. %
	The cut displayed in Figure \ref{sfig:bsp_algo2_5} is calculated using the FCE-\MCA with $C=\{e_{wy}, e_{yz}\}$, forcing $e_{wy}$ and $e_{yz}$ to be in the calculated cut.
	Note that the cut displayed in Figure \ref{sfig:bsp_algo2_5-counter} has a higher cut value, but because it does not cut edge $e_{wy}$ FCE-\MC{} will not consider it. %
	The cut displayed in Figure \ref{sfig:bsp_algo2_5} has the highest value of the three recursively calculated cuts, so it is transferred back to $H'$.
	The resulting cut is shown in Figure \ref{sfig:bsp_algo2_6}. 
	It is a maximum cut in $H'$.
\end{example}

\subsection{Analysis}

Building on the insights gained in Section \ref{se:alg-1pl}, we now are going to analyse Algorithm \ref{algo:MaxCut1pl-neg}. First we are going to transfer some propositions about $G-e$ to the adjusted graph $\AddEdges(G-e_{wz},\{e_{wy}, e_{yz}\})$. 
From Lemma \ref{lem:k-1-al-pl} in Section \ref{suse:alg-1pl:remCross} we can deduce Corollary \ref{cor:k-1-al-pl} and 
from Lemma \ref{lem:cut-transfer-from-G} (\ref{lem:cut-transfer-from-G_removal_wz}) in Section \ref{suse:alg-1pl:ana} we can deduce Corollary \ref{cor:cut-transfer-from-G_removal_wz}.

\begin{corollary}
	\label{cor:k-1-al-pl}
	Let $G$ be a $k$-almost-planar graph with a 1-planar embedding $(\I,\Pi)$ and let $\kz = \{e_{vy}, e_{wz}\} \in \I$ be an arbitrary crossing.
	The graph $\AddEdges$ $(G-e_{wz},\{e_{wy}, e_{yz}\})$ is $(k-1)$-almost-planar.
\end{corollary}

\begin{corollary}
	\label{cor:cut-transfer-from-G_removal_wz}	
	Let $G=(V,E,c)$ be a 1-planar graph with a 1-planar embedding $(\I,\Pi)$, $S\subseteq V$, and $\kz = \{e_{vy}, e_{wz}\} \in \I$ be an arbitrary crossing.
	Let $\Cut$ separate the four endpoints of $\kz$ as shown in Figure \ref{fig:8sepOptions} (d) or (h). Then $\nodeCutExt{_3} = \nodeCut$ defines a cut in $G_3=\AddEdges(G - e_{wz},\{e_{wy}, e_{yz}\})$ with the same value. If $\Cut$ is maximum in $G$, %
	so is $\CutExt{_3, G_3}$ in $G_3$ \dahn{under the condition that 
		$e_{wy}$ and $e_{yz}$
		are fixed cut edges}.
\end{corollary}

In lines \ref{algo:MaxCut1pl-neg:check1} and \ref{algo:MaxCut1pl-neg:check2}, the algorithm checks if merging two nodes would contradict the requirement that all edges in $C$ need to be in the calculated cut. 
Without this check the algorithm might run into inconsistencies because an edge cannot be forced in a cut after its nodes were merged.
In the following theorem we show that a cut calculated by Algorithm \ref{algo:MaxCut1pl-neg} is indeed maximum.

\begin{theorem}
	\label{theo:opt-new}
	Algorithm \ref{algo:MaxCut1pl-neg} computes a maximum cut in a weighted 1-~planar graph $G$ with arbitrary edge weights, given a set of crossing edges $X$ in a 1-planar embedding of $G$ and a set of fixed cut edges $C \subseteq E_G$.
\end{theorem}
\begin{proof}
	We prove its optimality by induction over $k=\vert X \vert$. Let $G$ be a weighted $k$-almost-planar graph.
	For $k=0$ the given graph is planar. Thus the \FCEMCA for planar graphs calculates a node set defining a maximum cut in $G$ \cite{LiersP12}.
	For $k>0$ we show that \dahn{the value of} the calculated cut is not smaller than \dahn{the value of} a maximum cut in $G$.
	Let $\nodeCutExt{^*}$ define a maximum cut in $G$ and let $\Cut$ be the cut defined by the calculated node set $\nodeCut$.
	Let $\kz = \{e_{vy}, e_{wz}\} \in \I$ be an arbitrary crossing. 
	Let $G_1 = G/wy, G_2 = G/yz$ %
	and $G_3 = \AddEdges(G - e_{wz},\{e_{wy}, e_{yz}\})$
	be the $(k-1)$-almost-planar graphs (Lemma \ref{lem:k-1-al-pl}, Corollary \ref{cor:k-1-al-pl}), whose cuts $\CutExt{_1, G_1}, \CutExt{_2, G_2}$ and $\CutExt{_3, G_3}$ were calculated recursively by the algorithm. 
	There are 8 possible ways for $\nodeCutExt{^*}$ to separate the four endpoints of $\kz$. These are shown in Figure \ref{fig:8sepOptions} (a)--(h).
	If the endpoints of $\kz$ are separated as shown in (a), (b), (c), (e), (f) or (g), the maximum of $\dahn{\wert(\CutExt{_1, G_1})}$ and $\dahn{\wert(\CutExt{_2, G_2})}$ is not smaller than $\dahn{\wert(\CutExt{^*, G})}$ (Theorem \ref{theo:opt}).
	If the endpoints of $\kz$ are separated as shown in (d) or (h), $\CutExt{^*, G}$ has the same value as a maximum cut $\CutExt{_3^*, G_3}$ in $G_3$ (Corollary \ref{cor:cut-transfer-from-G_removal_wz}). Due to the induction hypothesis, $\dahn{\wert(\CutExt{_3, G_3})}$ is not smaller than $\dahn{\wert(\CutExt{_3^*, G_3})}$. 
	The algorithm chooses the node set defining the \dahn{cut with the highest value} (line \ref{algo:MaxCut1pl-neg:maxGew}--\ref{algo:MaxCut1pl-neg:convert_E}) 
	and transfers it back to $G$ without changing its value (Lemma \ref{lem:cut-transfer-to-G}). Thus \dahn{the value of} the calculated cut $\Cut$ is not smaller than $\dahn{\wert(\CutExt{^*, G})}$ and $\Cut$ is therefore maximum in $G$.
\end{proof}

In order to gain an intuition about the running time of Algorithm~\ref{algo:MaxCut1pl-neg}, 
we consider its recursion tree. At each inner node a crossing is resolved by 
performing three recursive calls, leading to three children. Therefore, the 
height of the tree corresponds to the number of crossings and the leaves 
represent the planar problem instances. 
Assuming we have $k$ crossings, there are $3^k$ leaves, on which a planar problem
instance must be solved. The running time required for this dominates the running 
time required at the inner nodes. 
We now give a formal proof of the running time.

\begin{theorem}
	\label{theo:time}
	Let $G$ be an embedded edge-weighted 1-planar graph with $n$ nodes and $k$ crossings. %
	Algorithm \ref{algo:MaxCut1pl-neg} computes a maximum cut in $G$ in time $\mathcal{O}(3^{k} \cdot n^{3/2} \log n)$ if the planar \FCEMCA suggested in \cite{LiersP12} is used. 
\end{theorem}
\begin{proof}

	Let $T(k,n)$ be the running time of Algorithm \ref{algo:MaxCut1pl-neg} on an embedded edge-weighted 1-planar graph $G$ with $n$ nodes, $m$ edges and $k$ crossings in the given embedding. 
	Let $T_{p}(n)$ be the running time of the planar \FCEMC{} algorithm suggested by Liers and Pardella \cite{LiersP12}. 	
	If $G$ is planar, the running time is $T(0,n) = T_{p}(n)$.
	\\
	For $k>0$ choosing an arbitrary crossing %
	can be done in constant time. %
	$\contract$ has a linear running time of $\mathcal{O}(k)$, since every crossing in $\I$ needs to be checked only once.
	The contractions of $G/wy$ and $G/yz$ take time $\mathcal{O}(n + m)$ and the edge removal $G - e_{wz}$ takes time $\mathcal{O}(m)$. 
	$\AddEdges$ takes $\mathcal{O}(m)$ time because it needs to check if the edges already exist in the given graph, before adding them.
	For every crossing %
	two edges are added to $C$. Therefore $\vert C\vert$ is bound by $2k$.
	To check if two nodes, that are about to be merged, are connected by an edge in $C$ (which would contradict the requirement that all edges in $C$ need to be in the calculated cut) takes $\mathcal{O}(\vert C \vert) = \mathcal{O}(k)$ time.
	Reversing a contraction on a set of nodes $S_i$ with $\Split$ takes $\vert S_i\vert$ steps, resulting in a running time of $\mathcal{O}(n)$.
	\\
	Since $m$ is bound by $4n-8$ \cite{PachT97} and $k$ is bound by $2n-4$ (see above Theorem \ref{theo:time_calc_mc_kap}) the recursive running time is: 
	$T(k,n) = 3 \cdot T(k-1,n) + \mathcal{O}(k + n + m) = 3 \cdot T(k-1,n) + \mathcal{O}(n)$. 
	An induction proof confirms the closed form of the running time: $$T(k,n) = 3^k \cdot \left[T(0,n) + \sum_{i=1}^{k} 3^{-i} \cdot \mathcal{O}(n)\right]$$
	Since the geometric sum equals a value between 0 and 1, the overall running time is $\mathcal{O}(3^k \cdot (T_{p}(n) + n))$.
	The planar \FCEMCA has a running time of $T_{p}(n) = \mathcal{O}(n^{3/2} \log n)$ \cite{LiersP12}, leading to the concrete running time of $\mathcal{O}(3^k \cdot n^{3/2} \log n)$ for our algorithm. %
\end{proof}

In the above drawn picture of the recursion tree, $\mathcal{O}(n)$ would be the running time required at an inner node of the tree and $T_{p}(n) = \mathcal{O}(n^{3/2} \log n)$ would be the running time required at a leaf of the tree. The tree has at most $3^k$ leaves and less than $3^k$ inner nodes. Therefore the running time is $\mathcal{O}(3^k \cdot (T_{p}(n) + n))$. Since $T_{p}(n) = \mathcal{O}(n^{3/2} \log n)$ dominates $\mathcal{O}(n)$ the overall running time is $\mathcal{O}(3^{k} \cdot n^{3/2} \log n)$.

Regarding the parameterization of Algorithm \ref{algo:MaxCut1pl-neg}, we can deduce that it is \textit{fixed-parameter tractable} with parameter $k$ with the same arguments as in Section \ref{suse:alg-1pl:time}. This leads to the following conclusion about the weighted \MCP{} on embedded 1-planar graphs:

\begin{corollary}
	\label{cor:fpt-neg}
	The weighted \MCP\ on embedded 1-planar graphs with arbitrary edge weights and with $k$ crossings in the given embedding is \textit{fixed-parameter tractable} with parameter $k$.
\end{corollary}

\section{Conclusion and open problems}
\label{se:conclusion}
We have presented two polynomial time algorithms for computing a weighted \MC~in a 1-planar graph with non-negative resp.\@\xspace arbitrary edge weights given %
a 1-planar embedding with a constant number of crossings. If the number of crossings $k$ in the given embedding is not constant, we showed that the weighted \MCP on embedded 1-planar graphs is in the class FPT parameterized by $k$. 

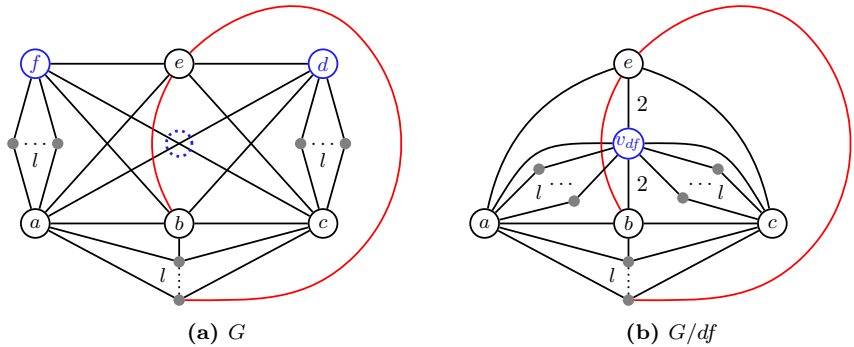
\begin{figure}
	\centering
	\begin{subfigure}{.49\linewidth}
		\centering
		\scalebox{.85}{%
\begin{tikzpicture}
\begin{scope} [thick] 
		\node[draw,circle,minimum size=0.45cm, inner sep=0ex] (a) at (0   ,0  ) {\small $a$};
		\node[draw,circle,minimum size=0.15cm,fill,gray, inner sep=0ex] (af1) at (-0.35 ,1.25) {};
		\node (afk) at (0   ,1.25) {\small $\ldots$};
		\node (afk) at (0 ,1) {\small $l$};
		\node[draw,circle,minimum size=0.15cm,fill,gray, inner sep=0ex] (afn) at (0.35 ,1.25) {};
		\node[draw,circle,minimum size=0.45cm, inner sep=0ex,blue!90] (f) at (0   ,2.5) {\small $f$};
		
		\node[draw,circle,minimum size=0.45cm, inner sep=0ex] (b) at (2.25,0  ) {\small $b$};
		\node[draw,circle,minimum size=0.45cm, inner sep=0ex] (e) at (2.25,2.5) {\small $e$};
		
		\node[draw,circle,minimum size=0.45cm, inner sep=0ex] (c) at (4.5 ,0  ) {\small $c$};		
		\node[draw,circle,minimum size=0.15cm,fill,gray, inner sep=0ex] (cd1) at (4.15 ,1.25) {};
		\node (cdk) at (4.5 ,1.25) {\small $\ldots$};
		\node (cdk) at (4.5 ,1) {\small $l$};
		\node[draw,circle,minimum size=0.15cm,fill,gray, inner sep=0ex] (cdn) at (4.85 ,1.25) {};
		\node[draw,circle,minimum size=0.45cm, inner sep=0ex,blue!90] (d) at (4.5 ,2.5) {\small $d$};

		\node[draw,circle,minimum size=0.15cm,fill,gray, inner sep=0ex] (ac1) at (2.25 ,-0.6) {};
		\node[draw,circle,minimum size=0.15cm,fill,gray, inner sep=0ex] (acn) at (2.25 ,-1.2) {};
		\node (ack) at (2,-0.8) {\small $l$};

	\draw (a) -- (d);
	\draw (a) -- (e);
	\draw (b) -- (d);
	\draw (b) -- (f);
	\draw (c) -- (e);
	\draw (c) -- (f);
	
	\draw[-,red] (b) to[bend left] (e);
	 	
	\draw (e) -- (f);
	\draw (e) -- (d);
	\draw (b) -- (a);
	\draw (b) -- (c);

	\draw (a) -- (af1) -- (f);
	\draw (a) -- (afn) -- (f);
	
	\draw (c) -- (cd1) -- (d);
	\draw (c) -- (cdn) -- (d);
	
	\draw (a) -- (ac1) -- (c);
	\draw (a) -- (acn) -- (c);
	
	\draw[-] (b) to (ac1);
	\draw[-,dotted] (ac1) to (acn);
	
	\draw[-,red] (e) to[out=45,in=135] (5,3) to[out=-45,in=45] (5,-0.5) to[out=-135,in = 0] (acn);
	
	\node[draw,circle,minimum size=0.4cm, inner sep=0ex,blue!90, dotted, very thick] (f) at (2.25,1.25) {};
\end{scope}
\end{tikzpicture} %
}	
		\caption{$G$}
		\label{fig:4-pl}
	\end{subfigure}
	\begin{subfigure}{.49\linewidth}
		\centering
		\scalebox{.85}{%
\begin{tikzpicture}
\begin{scope} [thick] 
		\node[draw,circle,minimum size=0.45cm, inner sep=0ex] (a) at (0   ,0  ) {\small $a$};
		\node[draw,circle,minimum size=0.15cm,fill,gray, inner sep=0ex] (af1) at (0.85 ,0.85) {};
		\node (afk) at (1.2 ,0.625) {\small $\ldots$};
		\node (afk) at (0.85 ,0.5) {\small $l$};
		\node[draw,circle,minimum size=0.15cm,fill,gray, inner sep=0ex] (afn) at (1.4 ,0.35) {};
		\node[draw,circle,minimum size=0.45cm, inner sep=0ex,blue!90] (vdf) at (2.25,1.25) {\small $v_{df}$};
		
		\node[draw,circle,minimum size=0.45cm, inner sep=0ex] (b) at (2.25,0  ) {\small $b$};
		\node[draw,circle,minimum size=0.45cm, inner sep=0ex] (e) at (2.25,2.5) {\small $e$};
		
		\node[draw,circle,minimum size=0.45cm, inner sep=0ex] (c) at (4.5 ,0  ) {\small $c$};		
		\node[draw,circle,minimum size=0.15cm,fill,gray, inner sep=0ex] (cd1) at (3.65 ,0.85) {};
		\node (cdk) at (3.375 ,0.625) {\small $\ldots$};
		\node (afk) at (3.7 ,0.5) {\small $l$};
		\node[draw,circle,minimum size=0.15cm,fill,gray, inner sep=0ex] (cdn) at (3.1 ,0.4) {};

		\node[draw,circle,minimum size=0.15cm,fill,gray, inner sep=0ex] (ac1) at (2.25 ,-0.6) {};
		\node (ack) at (2,-0.8) {\small $l$};
		\node[draw,circle,minimum size=0.15cm,fill,gray, inner sep=0ex] (acn) at (2.25 ,-1.2) {};

	\draw[-] (a) to[bend left, looseness=1.4] (vdf);
	\draw[-] (a) to[bend left] (e);
	\draw[-,thick] (b) to node[right] {2} (vdf);
	\draw[-] (c) to [bend right] (e);
	\draw[-] (c) to [bend right, looseness=1.4] (vdf);
	
	\draw[-,red] (b) to [bend left] (e);
	 	
	\draw[-,thick] (e) to node[right] {2} (vdf);
	\draw (b) -- (a);
	\draw (b) -- (c);

	\draw (a) -- (af1) -- (vdf);
	\draw (a) -- (afn) -- (vdf);
	
	\draw[-] (c) to (cd1) to (vdf);
	\draw[-] (c) to (cdn) to (vdf);
	
	\draw (a) -- (ac1) -- (c);
	\draw (a) -- (acn) -- (c);
	
	\draw[-] (b) to (ac1);
	\draw[-,dotted] (ac1) to (acn);
	
	\draw[-,red] (e) to[out=45,in=135] (5,3) to[out=-45,in=45] (5,-0.5) to[out=-135,in = 0] (acn);

\end{scope}
\end{tikzpicture} %
}
		\caption{$G/df$}
		\label{fig:4-pl_contr}
	\end{subfigure}
	\caption{A 4-planar graph where the contraction of the nodes $d$ and $f$ (colored blue) leads to $\mathcal{O}(l)$ new crossings. The two edges, that generate the new crossings, are drawn in red.
	Between $a$ and $f$, resp. $c$ and $d$, in $G$ are $l$ independent paths. Beneath $b$ there are $l$ paths between $a$ and $c$ that are pairwise connected and therefore have a specific order. The highest path contains a node connected to $b$ and the lowest path contains a node connected to $e$. No matter where $e$ is drawn in $G/df$, one of the two red edges crosses at least $l-1$ other edges.}
	\label{fig:k33}
\end{figure}

The question arises if our approach can be extended to general graphs with up to $k$ crossings per edge, so called \emph{$k$-planar graphs}.
Our approach is based on the fact that node contractions and edge deletions decrease the number of crossings. 
Figure \ref{fig:k33} shows that this is no more true if an edge is crossed more than once.
In this case there are crossings that do not have direct half edges connecting it to its endpoints like, e.g.,  the crossing $(ad,cf)$ in Figure \ref{fig:k33}. If we contract $d$ and $f$, we get plenty of new crossings in the new graph $G/df$.
\dahn{This shows that it is not trivial} to generalize our approach to embedded $k$-planar graphs.

Another interesting question would be to drop the assumption that we are given a 1-planar embedding. Note that our algorithm does not need such an embedding as input, it only needs to get a list of edge crossings. However, for our correctness analysis it is important to have a 1-planar embedding of the graph.

	In follow up works Chimani \etAl \cite{CDJKMN19} and Kobayashi \etAl \cite{KKMT19I} extended our approach 
	to general graphs and improved the running time to $\mathcal{O}(2^k \cdot (n+k)^{3/2} \log (n+k))$, where 
	$n$ is the number of nodes and $k$ is the number of crossings in a given embedded graph. 
	Furthermore, Chimani \etAl \cite{CDJKMN19} were able to drop the assumption of a given embedding to show 
	that \MC\ is fixed-parameter tractable with respect to the crossing number of the given graph. 
	Moreover, they showed that this result naturally carries over to the minor-monotone-version of the crossing number \cite{CDJKMN19}.
	
	It remains an open question whether the basis of the exponential dependency on $k$ can be reduced to less than $2$ or 
	if the running time of the planar \MCA\ and thus our overall running time can be improved further.

\bibliographystyle{plain}  %
\bibliography{literatur}

\clearpage

\end{document}